\documentclass[11pt]{article}
\usepackage{jheppub}

\usepackage{amsfonts,amssymb,amsmath,amsthm,url,graphicx,mathtools}
\usepackage{mathrsfs}
\usepackage{bbold}
\usepackage{datetime}
\usepackage{url}
\usepackage{multirow}
\usepackage{psfrag}

\numberwithin{equation}{section}

\DeclareMathOperator{\gl}{\mathfrak{gl}}
\DeclareMathOperator{\sgl}{\mathfrak{sl}}

\DeclareMathOperator{\U}{U}
\DeclareMathOperator{\SU}{SU}
\DeclareMathOperator{\so}{\mathfrak{so}}
\DeclareMathOperator{\osp}{\mathfrak{osp}}

\DeclareMathOperator{\su}{\mathfrak{su}}
\DeclareMathOperator{\uu}{\mathfrak{u}}

\DeclareMathOperator{\tr}{tr}

\newcommand{\D}{\mathcal{D}}
\newcommand{\y}{\hat{y}}
\newcommand{\KS}{{\mathrm{KS}}}

\newcommand{\shs}{\mathfrak{shs}}

\newcommand{\be}{\begin{equation}}
\newcommand{\ee}{\end{equation}}

\DeclareMathOperator{\Mat}{Mat}

\DeclareMathOperator{\ch}{ch}

\numberwithin{equation}{section}

\def\be{\begin{equation}}
\def\ee{\end{equation}}

\title{On the coset duals of extended higher spin theories}

\author{Constantin Candu and 
Carl Vollenweider}

\affiliation{
Institut f\"ur Theoretische Physik, ETH Z\"urich \\
CH-8093 Z\"urich, Switzerland}

\emailAdd{canduc@itp.phys.ethz.ch}
\emailAdd{carlv@itp.phys.ethz.ch}

\abstract{
We study the holographic duality between the $M\times M$ matrix extension of  Vasiliev higher spin theories on $AdS_3$ and the large $N$ limit of 
$\SU(N+M)/\SU(N)\times \U(1)$ type cosets.
We present a simplified proof for the agreement of the  spectra
and clarify the relation between this duality and the version in which the cosets are replaced by Kazama-Suzuki models of Grassmannian type.
}

\date{\today}

\allowdisplaybreaks

\begin{document}
\maketitle

\section{Introduction}\label{sec:intro}

Recently, some progress has been made on a concrete example in relating  higher spin/vector model holographic dualities to standard gauge/string theory dualities. 
In \cite{Chang:2012kt} it was proposed
that Vasiliev higher spin theories on $AdS_4$ with $\U(M)$ Chan-Paton factor are dual to the large $N$ limit of $\U(M)\times \U(N)$ ABJ gauge theories described by vector like CFTs.
Since the ABJ theories are believed to have a string theory dual \cite{Aharony:2008gk, Aharony:2008ug}, this suggests that the Vasiliev, ABJ and string theories are related by a triality.
The connection with string theory requires taking $M$ large, while the case  $M=1$ corresponds to the original higher spin/vector model duality of Klebanov \& Polyakov \cite{Klebanov:2002ja}, which was extensively tested in \cite{Giombi:2010vg, Giombi:2009wh}.

On the other hand, in one dimension lower, the minimal model holography \cite{Gaberdiel:2010pz, Gaberdiel:2012ku} (see  \cite{Gaberdiel:2012uj} for a review) and its generalizations 
\cite{Ahn:2011pv, Gaberdiel:2011nt, Creutzig:2011fe, Creutzig:2012ar, Beccaria:2013wqa} have convincingly shown that higher spin/vector model dualities are much simpler to deal with.
Thus, if one can find a generalization of the triality~\cite{Chang:2012kt} in this context,  then one  naturally expects to be able to understand in great detail the mechanism by which the CFT, the Vasiliev theory and the string theory are connected to each other.

From this perspective,  dualities between
$M\times M$ matrix extensions of the $\mathcal{N}=2$ Vasiliev higher spin theories on $AdS_3$ \cite{Prokushkin:1998bq} and the large $N$ limit 
of  $\mathcal{N}=1$ cosets of the form $\SU(N+M)/\SU(N)\times \U(1)$ have been proposed in \cite{Gaberdiel:2013vva, Creutzig:2013tja}. 
These are in many respects analogous to the above mentioned dualities between the $\U(M)$
extended Vasiliev theories and the large $N$ limit of $\U(M)\times \U(N)$ ABJ theories.
A convenient way to think about them is as ``matrix generalizations'' of the $\mathcal{N}=2$ example \cite{Creutzig:2011fe} (see also \cite{Candu:2012jq})
of the minimal model holography 
to which they reduce for $M=1$.

According to the crucial observation of \cite{Gaberdiel:2013vva}, the supersymmetry of the extended Vasiliev theory is enhanced from $\mathcal{N}=2$ to large
$\mathcal{N}=4$ when $M$ is even or, more precisely,  the higher spin algebra 
  of the extended  Vasiliev theory $\shs_M[\mu]$  contains the large $\mathcal{N}=4$ supersymmetry algebra $D(2,1;\alpha)$ with $\alpha = \mu/(1-\mu)$ as a subalgebra for any even $M$.
This finding is of particular interest since it is believed that there is essentially only one string background which supports this 
supersymmetry, namely $AdS_3 \times S^3 \times S^3 \times S^1$ \cite{deBoer:1999rh, Gukov:2004ym}.\footnote{Currently, the CFT dual of this string theory is not known. See \cite{Gukov:2004ym} for attempts to find it.}
And indeed, it was checked in \cite{Gaberdiel:2013vva} that the (multi-particle) BPS states of the $M=2$ Vasiliev theory reproduce  the spectrum of  single particle BPS states of supergravity on $AdS_3 \times S^3 \times S^3 \times S^1$.
Based on this result and the natural expectation that for large $M$ one can generate the full (multi-particle) BPS spectrum of supergravity, the authors of \cite{Gaberdiel:2013vva} have concluded that the extended Vasiliev theory has a good chance of being dual to string theory on $AdS_3 \times S^3 \times S^3 \times S^1$.
On the other hand, it was also observed that the cosets do not have  large $\mathcal{N}=4$ supersymmetry, except for the special case of $M=2$ which belongs to the classification of \cite{Spindel:1988sr, VanProeyen:1989me, sevrin} based on Wolf spaces.
In contrast, the case $M=2$ did not receive any special attention in the reference \cite{Creutzig:2013tja}, in which  
the 1-loop partition function of the extended Vasiliev theory was matched with the 't~Hooft limit of the coset partition function for general $M$.

In this work we shall study the duality of \cite{Gaberdiel:2013vva, Creutzig:2013tja} for general $M$ in more detail.
Our aim is twofold.  
First, we carry out a simplified analysis of the agreement between the 1-loop partition function of the Vasiliev theory and the 't~Hooft limit of the coset partition function.
In particular, we include chemical potentials for the residual affine symmetries which, in principle, can be used to extract the  BPS spectrum.

Our second goal is to elucidate the relation between the above duality and its slight modification 
in which the cosets are of the form
$\SU(N+M)/\SU(N) \times \SU(M)\times  \U(1)$, i.e.\ Grassmannian Kazama-Suzuki type coset. In fact, this point has lead to some confusion in \cite{Creutzig:2013tja} which we would like to clarify here.
To this end, we first show that in the 't~Hooft limit the Kazama-Suzuki cosets are equivalent to the previous $\SU(N+M)/\SU(N)\times \U(1)$ cosets plus constraints.
The constraints restrict only the 
currents  and we discuss at length their effect on the partition function and the $\mathcal{W}$-algebra of the theory.
The main point, however, is that these constraints can be implemented on the higher spin side simply by refining the standard asymptotic AdS boundary conditions without changing the asymptotic AdS geometry.
Thus, it is the boundary conditions of the Vasiliev theory that determine the precise form of the coset dual.

The paper is organized as follows. In section 2 we introduce the matrix extension of the $\mathcal{N}=2$ Vasiliev theory and  briefly describe its gauge algebra, field content, dynamics, boundary conditions, asymptotic symmetry algebra and 1-loop partition function.
In section 3 we introduce the proposed coset duals, compute their  partition function and higher spin spectrum in the 't~Hooft limit and check the agreement with predictions from the higher spin theory.
In section 4 we discuss the modified duality in which the cosets are replaced by Kazama-Suzuki models, while  the Vasiliev theory is subject to slightly modified asymptotic AdS boundary conditions. 
Finally, in section 5 we conclude and comment on the supersymmetry problem for $M>2$.


\section{Extended higher spin theories}\label{sec:extended}

The aim of this section is to introduce the matrix extension of the $\mathcal{N}=2$ higher spin theory of $AdS_3$ gravity of Prokushkin and Vasiliev \cite{Prokushkin:1998bq}. 
The higher spin theory being a  gauge theory, we shall first describe its gauge algebra.
We then proceed to describe the fields of the theory, their dynamics, the emergence of asymptotic symmetries and the 1-loop partition function.

\subsection{Extended higher spin algebra}

Consider the following associative algebra
\begin{equation}
sB[\mu] = U(\osp(1|2))/\langle \mathrm{Cas}-\tfrac{1}{4}\mu(\mu-1)\mathbb{1}\rangle \simeq sB[1-\mu]
\label{eq:sB}
\end{equation}
obtained as a quotient of the universal enveloping algebra of the Lie superalgebra $\osp(1|2)$ by a central ideal.
In the standard $\mathcal{N}=1$ superconformal like basis of $\osp(1|2)$
\begin{equation}
[L_m,L_n]=(m-n)L_{m+n}\ ,\quad [L_m,G_r] = (m/2-r)G_{r+m}\ ,\quad \{G_r,G_s\} = 2 L_{r+s}\ ,
\label{eq:osp12}
\end{equation}
where $m,n=-1,0,1$ and $r,s=\pm 1/2$, the Casimir is normalized as
\begin{equation}
\mathrm{Cas} = L_0^2 - \tfrac{1}{2}\{L_1,L_{-1}\}+\tfrac{1}{4}[G_{1/2},G_{-1/2}]\ .
\label{eq:casosp12}
\end{equation}
The algebra $sB[\mu]$ can be faithfully realized in terms of two oscillators $\y_1$, $\y_2$ and a Kleinian operator $k$ satisfying the relations
\begin{equation}
[\y_\alpha,\y_\beta]=\y_\alpha\y_\beta-\y_\beta\y_\alpha=2i\epsilon_{\alpha\beta}(\mathbb{1}+\nu k)\ ,\quad k \y_\alpha = - \y_\alpha k\ ,\quad k^2 = \mathbb{1}\ ,
\label{eq:oscalg}
\end{equation}
where $\epsilon_{\alpha\beta}= - \epsilon_{\beta\alpha}$ and $\epsilon_{12}=1$, $\nu=2\mu-1$ if we identify
\begin{equation}\label{eq:gpmh}
G_{\frac{1}{2}} = \tfrac{1}{2}e^{-i\frac{\pi}{4}}\y_1\ ,\quad  G_{-\frac{1}{2}} = \tfrac{1}{2}e^{-i\frac{\pi}{4}}\y_2\ ,
\end{equation}
and
\begin{equation}
L_1= \tfrac{1}{4i}\y_1^2\ ,\quad L_{-1}= \tfrac{1}{4i}\y_2^2\ , \quad L_0 =\tfrac{1}{8i}(\y_1\y_2+\y_2\y_1) \ .
\label{eq:emt}
\end{equation}
In other words, the oscillator algebra  generated by $\y_\alpha$ and $k$ is isomorphic to $sB[\mu]$ if $\nu=2\mu-1$.
To see this, one first checks that the generators~(\ref{eq:gpmh}, \ref{eq:emt})
satisfy the commutation relations~\eqref{eq:osp12} and that $\mathrm{Cas} =(\nu^2 k^2  - 1)/16 = \mu(\mu-1)/4$.
Second, one must check that the vector space generated by the products of $\y_\alpha$ and $k$ has the same ``dimension'' as $sB[\mu]$.
For this it is sufficient to prove  that both of these spaces decompose as
\begin{equation}
\bigoplus_{j=0}^\infty 2\times D_j\ ,
\label{eq:dec_sl2}
\end{equation}
where the sum is over all representations $D_j$ of $\mathfrak{sl}(2)\subset \osp(1|2)$ of half-integer spin $j$.
This decomposition is equivalent to the statement that both $sB[\mu]$ and the oscillator algebra have exactly two $\mathfrak{sl}(2)$ highest weight states at every half-integer spin $j\geq 0$.
Clearly,  for $sB[\mu]$ these are $G^{2j}_{1/2}$ and
$ [G_{1/2}, G_{-1/2}]G^{2j}_{1/2}$, while for the oscillator algebra these are $\y_1^{2j}$ and $k \y_1^{2j}$, where we have used the relation $G_{1/2}^2 = L_1$ and the fact that $[G_{1/2}, G_{-1/2}]$ is an $\mathfrak{sl}(2)$ singlet.



The structure constants of the multiplication operation in $sB[\mu]$ are known explicitly~\cite{Pope:1989sr}. These are usually presented in a basis of symmetrized  products of oscillators
\begin{equation} \label{basissB}
V_m^{(s)\pm} \propto \y_{(\alpha_1...}\, \y_{\alpha_{l})}P_\pm\ , 
\end{equation}
where $P_\pm=(\mathbb{1} \pm k)/2$ are projectors, $s=\frac{l}{2}+1$, and  $2m=\#\y_1-\#\y_2$ takes values in the range
$-s+1\leq m \leq s-1$. 
Notice that the  basis vectors $V_m^{(s)\pm}$ with $|m|<s$ span the representations $D_{s-1}$ in eq.~\eqref{eq:dec_sl2}.
Due to the commutation relations 
\begin{equation}
[L_m,V^{(s)\pm}_n] = [m(s-1)-n]V^{(s)\pm}_{m+n}
\end{equation}
they are called  generators of conformal spin $s$.

Let us define a $\mathbb{Z}_2$ grading $|\cdot|$ on $sB[\mu]$ by calling the generators $V^{(s)\pm}_m$ even or bosonic if  $s$ is integer and odd or fermionic if $s$ is half odd-integer.
Then one can turn $sB[\mu]$   into a Lie superalgebra by endowing it with the usual Lie bracket $[a,b]_\pm := ab - (-1)^{|a||b|}ba$.
Moreover, according to \cite{Vasiliev:1989re, Fradkin:1990qk}  $sB[\mu]$ has a graded symmetric trace which is non-degenerate for $\mu\notin\mathbb{Z}$.
Thus, the traceless part of $sB[\mu]$  will form a subalgebra $\shs[\mu]$ which is simple for $\mu\notin\mathbb{Z}$.\footnote{
For $\mu\in\mathbb{Z}$, $\shs[\mu]$ acquires a unique maximal ideal $\chi_\mu$ such that the quotient $\shs[\mu]/\chi_\mu$ is simple and isomorphic to $\sgl(\mu|\mu-1)$ if $\mu>0$ and $\sgl(1-\mu|-\mu)$ if $\mu<0$, see 
\cite{Fradkin:1990qk}.}


Let us now extend the associative algebra $sB[\mu]$ by tensoring it with the  matrix algebra $\Mat_M$ of complex $M\times M$ matrices 
\begin{equation}
sB[\mu]_M := sB[\mu]\otimes \Mat_M\ ,
\label{eq:extsB}
\end{equation}
which can also be viewed as the associative algebra of $sB[\mu]$ valued $M\times M$ matrices. It inherits a natural conformal  $\sgl(2)$ subalgebra
\begin{equation}\label{eq:sl2ext}
L_m \equiv L_ m  \otimes \mathbb{1}_M\ ,
\end{equation}
a parity grading $|a\otimes A|= |a|$, and a graded symmetric trace
\begin{equation}\label{def_trace}
\tr a\otimes A = \tr a \tr A\ .
\end{equation}
Hence, $sB_M[\mu]$ can be turned into a Lie superalgebra in the usual way. Its traceless part will form a subalgebra $\shs_M[\mu]$
which is simple for $\mu\notin \mathbb{Z}$ and which decomposes as follows 
\begin{equation}
\shs_M[\mu] := \mathbb{1}\otimes \mathfrak{sl}(M)\; \oplus\; \shs[\mu]\otimes \mathbb{1}_M\;\oplus\;\shs[\mu]\otimes \mathfrak{sl}(M)
\label{eq:extshs}
\end{equation}
w.r.t.\ the action of the subalgebra $\shs[\mu]\otimes \mathbb{1}_M\oplus \mathbb{1}\otimes \sgl(M)$.
The  spin $s=1$ subspace of $\shs[\mu]$ is the direct sum of the two  mutually commuting subalgebras 
\begin{equation}
\mathfrak{sl}(M)_\pm:=P_\pm\otimes \mathfrak{sl}(M)
\label{eq:slnpm}
\end{equation}
together with the $\gl(1)$ subalgebra generated by $J_0\otimes \mathbb{1}_M$, where 
\begin{equation}
J_0 := -\tfrac{1}{2}(\nu\mathbb{1}+k)
\label{eq:j0}
\end{equation}
spans the traceless elements of $\shs[\mu]$ ar $s=1$.
 Using the explicit basis \eqref{basissB}, it is easy to check that the spin $s$ part of $\mathfrak{shs}_M[\mu]$ decomposes into the following multiplets of $\mathfrak{sl}(M)_+\oplus \mathfrak{sl}(M)_-\oplus \gl(1)$
\begin{align}\label{eq:dec_shsn}
s=1:&\quad  (adj, 0)_0 \oplus(0,adj)_0\oplus (0,0)_0\\ \notag
s\in \mathbb{N}+\tfrac{1}{2}:&\quad (f,f^*)_{-1}\oplus(f^*,f)_{1}\\ \notag
s\in\mathbb{N}+1:&\quad (adj,0)_0\oplus(0,adj)_0\oplus 2 (0,0)_0\ ,
\end{align}
where $f$ is the fundamental representation of $\mathfrak{sl}(M)$, $f^*$ is its dual, $adj$ is the adjoint representation and the index denotes the $J_0$ charge.

To construct the Vasiliev theory, one needs to impose a reality condition on $\shs_M[\mu]$.
In the following we shall assume that this reality condition selects the unitary real forms of the  subalgebras  $\sgl(M)_\pm$ at spin $s=1$.

\subsection{Extended Vasiliev theory}\label{sec:matV}


The explicit form of the e.o.m.\ of the extended Vasiliev higher spin theory based on $\shs_M[\mu]$ can be found in \cite{Prokushkin:1998bq}.
For our purposes it suffices to consider the simpler form of these equations, which is linear in the matter fields \cite{Prokushkin:1998bq} (see also \cite{Ammon:2011ua})
\begin{align}\label{eq:leom}
dA + A \wedge A&=0\ ,&  d\bar A + \bar A \wedge \bar A&=0\ ,\\
d C + A  C - C  \bar{A} &= 0 \ ,&d \bar C + \bar A  \bar C - \bar C  A &= 0 \ ,\label{eq:mattereom}
\end{align}
where  $A$, $\bar A$ are  1-forms taking  values in $\shs_M[\mu]$ and $C$, $\bar C$ are 0-forms taking values in $sB_M[\mu]$.
These equations are invariant w.r.t.\ the gauge transformations
\begin{align}\label{eq:gtra}
\delta_{\Lambda,\bar{\Lambda}} A&= d\Lambda +[A,\Lambda]\ ,&\delta_{\Lambda,\bar{\Lambda}} \bar A&= d\bar \Lambda +[\bar A,\bar \Lambda]\ ,\\
\delta_{\Lambda,\bar{\Lambda}} C &=   C\bar \Lambda - \Lambda C
\ ,&\delta_{\Lambda,\bar{\Lambda}} \bar C &=   \bar C \Lambda - \bar \Lambda \bar C\ ,
\label{eq:gtrc}
\end{align}
where $\Lambda,\bar \Lambda\in \shs_M[\mu]$.
All fields  are defined on a 3-manifold  with the topology of a solid cylinder. Their dynamics can be roughly understood as follows.
If the connections $A$ and $\bar A$ were to take values in $\sgl(2)$, then the flatness conditions would be   equivalent to the vacuum Einstein e.o.m.\ 
with negative cosmological constant
in the first order formalism, see \cite{Achucarro:1987vz, Witten:1988hc}. The equivalence is established by expressing $A$ and $\bar A$ in terms of the vielbein $e$ and the spin connection $\omega$ as follows
\begin{equation}
A = \omega + \frac{e}{\ell}\ ,\qquad \bar A = \omega - \frac{e}{\ell}\ ,
\end{equation}
where $\ell$ is a length unit which can be identified with the radius of the globally $AdS_3$ solution
\begin{align}\notag
A_{AdS} &= e^{-L_0 \rho}\left(L_1+\frac{L_{-1}}{4}\right) e^{L_0 \rho} dx + L_0 d\rho\ ,\\
  \bar A_{AdS}&= - e^{L_0 \rho}\left(L_{-1}+\frac{L_{1}}{4}\right) e^{-L_0 \rho} d\bar x - L_0 d\rho\ .
\label{eq:absbckr}
\end{align}
Here $x = t/\ell+ \theta$, $\bar x = t/\ell- \theta$, $t$ is the time coordinate, $\theta$ is the angular coordinate and $\rho$ is the radial coordinate on the cylinder, with the boundary being located at  $\rho\to\infty$.

When the gauge fields $A$, $\bar A$  take values in $\shs_M[\mu]$, the flatness conditions can be understood as the e.o.m.\ in the first order formalism for the infinite tower \eqref{eq:dec_shsn} of gauge fields of all  half-integer spins $s\geq 1$;  for details on the first order formalism for higher spin gauge fields see \cite{Vasiliev:2003ev}.
In particular, the components of $A$ ($\bar A$) along the spin $s=1$ generators of $\shs[\mu]$ describe a set of (topological) vector fields valued in a left (right) copy of the subalgebra $\su(M)_+\oplus\su(M)_-\oplus\uu(1)$,
while the components along the $\sgl(2)$ generators~\eqref{eq:emt} describe as before the gravity field. More generally, the components of $A$ and $\bar A$  along the spin $s\geq \frac{3}{2}$ generators correspond to $2M^2$  spin $s$ gauge fields  which are  charged under the vector fields as described in eq.~\eqref{eq:dec_shsn}.
The reformulation of  these fields in the second order formalism requires introducing a notion of geometry, which comes in through the  asymptotically $AdS_3$ boundary condition, see \cite{Campoleoni:2010zq, Campoleoni:2012hp}
\begin{equation}
A - A_{AdS_3} \sim \mathcal{O}(\rho^0)\ ,\quad \bar A - \bar A_{AdS_3} \sim \mathcal{O}(\rho^0)\ , \qquad \qquad \rho\to \infty\ .
\label{eq:bc}
\end{equation}

The scalar fields $C$, $\bar C$ behave rather differently. Most of their components are non-dynamical and their only role is to provide a manifest representation of the higher spin algebra $\shs_M[\mu]$.
On the asymptotically $AdS_3$ higher spin background~\eqref{eq:bc}, one can use the e.o.m.~\eqref{eq:mattereom} as explained in \cite{Prokushkin:1998bq} to express all  components  only in terms of the lowest spin $s=1$ bosonic components and the lowest spin $s=\frac{3}{2}$ fermionic components, which are denoted by 
\begin{align}\notag
C &= P_+\otimes \phi_+ + P_-\otimes \phi_- +\hat y_\alpha P_+ \otimes \psi^\alpha_+ + \hat y_\alpha P_-\otimes \psi^\alpha_-+\mathcal{O}(\hat y^2)\ ,\\
 \bar C &= P_+\otimes \bar \phi_+ + P_-\otimes\bar  \phi_- +\hat y_\alpha P_+ \otimes \bar \psi^\alpha_+ + \hat y_\alpha P_-\otimes \bar \psi^\alpha_-+\mathcal{O}(\hat y^2)\ , 
\label{dyn_f}
\end{align}
where $\phi_\pm$, $\bar \phi_+$  are $\mathrm{Mat}_M$ valued bosonic complex scalar fields and $\psi_\pm$, $\bar \psi_\pm$ are $\mathrm{Mat}_M$ valued fermionic Dirac fields.
After this ``folding'' procedure the e.o.m.~\eqref{eq:mattereom} reduce to linear higher derivative e.o.m\ for the lowest spin components, describing their free propagation on the higher spin background (see \cite{Ammon:2011ua} for explicit examples of such equations).
In particular, on the pure $AdS_3$ background \eqref{eq:absbckr} the e.o.m.\ for $\phi_\pm$, $\bar \phi_\pm$ reduce to the Klein-Gordon equation on $AdS_3$ with mass
squared $M^2_\pm = -1+ (1\mp \nu)^2/4$, while the e.o.m.\ for $\psi^\alpha_\pm$, $\bar \psi^\alpha_\pm$ reduce to the Dirac equation on $AdS_3$ with mass squared 
squared $m^2_\pm=\nu^2/4$, where the masses are given in units of $\ell$.
The charges of  matter fields w.r.t.\ the vector fields of the theory can be easily derived from eqs.~(\ref{eq:mattereom}, \ref{dyn_f})
and are represented in table~\ref{tab}.
Notice that when the complex conjugate fields are taken into account  the quantum numbers of all fields are such that a degeneracy of 2 survives: $\phi_\pm$ is indistinguishable from $\bar \phi_\pm^*$ and $\psi_\pm$ from $\bar\psi_\mp^*$. To later match with the CFT, we have associated opposite quantizations to the  degenerate pairs, i.e.\ the conformal dimension $\Delta:=h+\bar h$ lifts the degeneracy.

\begin{table}%
\begin{center}
\begin{tabular}{|c|c|c|c|}\hline
Fields & Mass & Dimensions $\Delta$& Charges\\\hline
$\phi_+$ & $M^2_+=-1+(1-\mu)^2$ & $\mu$ & $(f, 0)_{-\mu}\otimes (f^*,0)_{\mu}$ \\ \hline
$\psi_+$ & \multirow{2}{*}{$m^2_\pm = (\mu-\frac{1}{2})^2$} & \multirow{2}{*}{$\mu+\frac{1}{2}$} & $(0,f)_{1-\mu}\otimes (f^*,0)_\mu$\\ \cline{1-1}\cline{4-4}
$\psi_-$ & & & $(f,0)_{-\mu}\otimes (0,f^*)_{\mu-1}$  \\ \hline
$\phi_-$ &$M^2_-=-1+\mu^2$ &  $\mu+1$& $(0,f)_{1-\mu}\otimes (0,f^*)_{\mu-1}$ \\ \hline \hline
$\bar \phi_-$ & $M^2_+=-1+\mu^2$ & $1-\mu$ & $(0,f^*)_{\mu-1}\otimes (0,f)_{1-\mu}$ \\ \hline
$\bar\psi_-$ & \multirow{2}{*}{$m^2_\pm = (\mu-\frac{1}{2})^2$} & \multirow{2}{*}{$\frac{3}{2}-\mu$} & $(0,f^*)_{\mu-1}\otimes (f,0)_{-\mu}$\\ \cline{1-1}\cline{4-4}
$\bar \psi_+$ &  & & $(f^*,0)_\mu\otimes (0,f)_{1-\mu}$ \\ \hline
$\bar \phi_+$ & $M^2_+=-1+(1-\mu)^2$ & $2-\mu$ & $(f^*, 0)_{\mu}\otimes (f,0)_{-\mu}$\\ \hline
\end{tabular}
\caption{The matter fields of the $\shs_M[\mu]$ Vasiliev theory. The left (right) factor in the charge denotes the transformation properties w.r.t.\ the $\su(M)_+\oplus \su(M)_-\oplus \uu(1)$ fields of $A$ ($\bar A$) at spin $s=1$ and the conventions for the various representation labels are the same as in eq.~\eqref{eq:dec_shsn}.}
\label{tab}
\end{center}
\end{table}

\subsection{Asymptotic symmetry algebra}\label{sec:asa}

According to \cite{Campoleoni:2010zq}, the most general flat connections $A$ and $\bar A$ satisfying the asymptotic boundary condition~\eqref{eq:absbckr}
with any remaining gauge freedom removed
can be written as
\begin{equation}
A = e^{-L_0 \rho}a e^{L_0 \rho}dx  + L_0 d\rho\ , \qquad \bar A=  -e^{L_0 \rho}\bar a e^{-L_0 \rho}d\bar x - L_0 d\rho\ ,
\label{eq:gg_tr}
\end{equation}
 where $a$ depends only on $x$, $\bar a$ only on $\bar x$ and they are of the form
\begin{align}\notag
a &= L_1 \otimes \mathbb{1}_M + P_+ \otimes t^I J^I + P_- \otimes t^I K^I+J_0\otimes \mathbb{1}_M U+\sum_{s\geq \frac{3}{2},\varepsilon=\pm } V^{(s) \varepsilon}_{-s+1} \otimes E_{ij} W^{(s) \varepsilon}_{ij} \ ,\\
\bar a &= L_{-1} \otimes \mathbb{1}_M + P_+ \otimes t^I \bar J^I + P_- \otimes t^I \bar K^I+J_0\otimes \mathbb{1}_M \bar  U+\sum_{s\geq \frac{3}{2},\varepsilon=\pm } V^{(s) \varepsilon}_{s-1} \otimes E_{ij} \bar W^{(s) \varepsilon}_{ij} \ .
\label{eq:gfix}
\end{align}
Here  $t^I$ is a basis of $\su(M)$ and $E_{ij}$ is the matrix with entry 1 at position $(ij)$ and zero everywhere else.
Hence,  the most general asymptotically AdS background is parametrized by two copies of left and right moving $2\,M^2-1$ spin $s=1$ currents, and $2\,M^2$ currents of every half-integer spin $s \geq \frac{3}{2}$.

The flatness conditions~\eqref{eq:leom} can be derived from a double  Chern-Simons theory for the Lie superalgebra $\shs_M[\mu]$ endowed with the trace~\eqref{def_trace}.
The latter comes with a canonical Poisson bracket which can be used to define a left moving Poisson-bracket algebra satisfied by the holomorphic currents~\eqref{eq:gfix}
together with a similar right moving copy.
As explained in great detail in \cite{Campoleoni:2010zq, Gaberdiel:2011wb}, these are generically non-linear $\mathcal{W}$-algebras which can be computed as the
Drinfel'd-Sokolov reduction of the gauge algebra of the Chern-Simons theory, in our case $\shs_M[\mu]$,  w.r.t.\ the $\sgl(2)$ subalgebra entering in the asymptotic boundary conditions~\eqref{eq:bc}.
They are called asymptotic symmetry algebras, because they act on a given classical solution~\eqref{eq:gfix} of the Chern-Simons theory as field dependent gauge transformations~\eqref{eq:gtra} that do not decay at the boundary and, for this reason, map the given solution to physically inequivalent solutions.

\subsection{Partition function}

\label{sec:hs}

One of the simplest characteristics of the perturbatively quantized Vasiliev theory is the 1-loop partition function on thermal $AdS_3$,
which is topologically a solid torus parametrized by the modular parameter $q=e^{2\pi i\tau}$ of the conformal boundary torus $\mathbb{C}/(\mathbb{Z}+\tau \mathbb{Z})$. 
The 1-loop partition function takes into account only the quadratic fluctuations of the fields around the Euclidean AdS vacuum, hence is uniquely determined by the field content of the theory
in terms of the following elementary building blocks: the partition function of an integer spin $s$ gauge field \cite{Giombi:2008vd, Gaberdiel:2010ar}
\begin{equation}\label{prob1}
\prod_{n=s}^\infty\frac{1}{(1-q^n)(1-\bar q^n)}\ ,
\end{equation}
the partition function of a half odd-integer (anti-periodic) spin $s$ gauge field  \cite{David:2009xg, Creutzig:2011fe}
\begin{equation}\label{prob2}
\prod_{n=s-\frac{1}{2}}^{\infty}(1+q^{n+\frac{1}{2}})(1+\bar q^{n+\frac{1}{2}})\ ,
\end{equation}
the partition function of a complex massive scalar of conformal dimension $\Delta=2h $ \cite{Giombi:2008vd}
\begin{equation}
\prod_{m,n=0}^\infty\frac{1}{(1-q^{h+m}\bar q^{h+n})^2}
\label{prob3}
\end{equation}
and, finally, the partition function of a (anti-periodic) massive Dirac fermion of conformal dimension $\Delta = 2h+\frac{1}{2}$ \cite{Creutzig:2011fe}
\begin{equation}\label{prob4}
\prod_{m,n=0}^\infty(1+q^{h+\frac{1}{2}+m}\bar q^{h+n})(1+q^{h+m}\bar q^{h+\frac{1}{2}+n})\ .
\end{equation}

In order to write down the 1-loop partition function explicitly, let us split it into a gauge part and matter part
\begin{equation}
Z^{\text{1-loop}}_{\text{Vasiliev}} = Z_{\text{gauge}} Z_{\text{matter}}\ .
\label{eq:totoal_pf}
\end{equation}
Then, with the help of eqs.~(\ref{prob1}, \ref{prob2}) and the decomposition~\eqref{eq:dec_shsn} we can write the gauge field contribution as follows
\begin{align}\notag
Z_{\text{gauge}} = 
\prod_{s=1}^\infty &\prod_{n=s}^\infty \prod_{i,j=1}^{M} \frac{(1+q^{n+\frac{1}{2}}z^i_{+} z_{-}^{j*})(1+q^{n+\frac{1}{2}}z^{i*}_{+} z_{-}^{j})}{(1-q^n z_+^i z_+^{j*})(1-q^n z_-^i {z_-^{j*}})}\frac{(1+\bar{q}^{n+\frac{1}{2}}\bar{z}^i_{+} \bar{z}_{-}^{j*})(1+\bar{q}^{n+\frac{1}{2}}\bar{z}^{i*}_{+} \bar{z}_{-}^{j})}{(1-\bar{q}^n \bar{z}_+^i \bar{z}_+^{j*})(1-\bar{q}^n \bar{z}_-^i \bar{z}_-^{j*})}\times \\
\times &\prod_{n=1}^\infty (1-q^n)(1-\bar q^n) \ ,
\label{eq:gauge_pf_shsn}
\end{align}
where we have introduced the phases $z_\pm^i$  ($\bar{z}_\pm^i$) ``by hand'' to keep track of the charges w.r.t.\ the left (right) $\mathfrak{su}(M)_\pm$ vector fields.
Heuristically this partition function can be understood as follows: every positive  Fourier mode of the spin $s$ boundary current $W^{(s)\pm }_{ij}(x)$ contributes  with a factor $q^n z^i_\pm z^{*j}_\pm$ if $s\in\mathbb{N}$ and $q^{n+\frac{1}{2}} z^i_\mp z^{*j}_\pm$ if $s\in\mathbb{N}+\frac{1}{2}$;  the modes with $n\leq s$ are discarded because they  correspond to residual symmetries of the AdS vacuum \cite{Perlmutter:2012ds, Hikida:2012eu}. The right current $\bar W^{(s)\pm }_{ij}(\bar x)$ contributes similarly to the right part.
%

Let us now split the matter part of the partition function
\begin{equation}\label{matter}
Z_{\text{matter}} = Z^{+}_{\text{matter}}Z^{-}_{\text{matter}}\ ,
\end{equation}
into a factor  $Z^{+}_{\text{matter}}$ containing the contribution of the fields $\phi_\pm$, $\psi_\pm$
and a factor  $Z^{-}_{\text{matter}}$ containing the contribution of $\bar\phi_\pm$, $\bar\psi_\pm$.
Defining
\begin{equation}
h_+ = \frac{\mu}{2}\ ,\qquad h_-=\frac{1-\mu}{2}\ ,
\end{equation}
and using eqs.~(\ref{prob3}, \ref{prob4}) together with table \ref{tab} we can write these factors as
\begin{multline}
Z^{\pm}_{\text{matter}} = \prod_{m,n=0}^\infty \prod_{i,j=1}^{M}\frac{(1+q^{h_\pm +m}\bar{q}^{h_\pm + \frac{1}{2}+n} z_\pm^i \bar{z}_\mp ^{*j})(1+q^{h_\pm +m}\bar{q}^{h_\pm + \frac{1}{2}+n} z_\pm^{*i} \bar{z}_\mp ^{j})}{(1-q^{h_\pm+m}\bar{q}^{h_\pm+n}z_\pm^i \bar{z}_\pm^{*j})(1-q^{h_\pm+m}\bar{q}^{h_\pm+n}z_\pm^{*i} \bar{z}_\pm^{j})}
\times \\
\times 
\frac{(1+q^{h_\pm+\frac{1}{2} +m}\bar{q}^{h_\pm + n} z_\mp^i \bar{z}_\pm ^{*j})(1+q^{h_\pm+\frac{1}{2} +m}\bar{q}^{h_\pm + n} z_\mp^{*i} \bar{z}_\pm ^{j})}{(1-q^{h_\pm+\frac{1}{2}+m}\bar{q}^{h_\pm+\frac{1}{2}+n}z_\mp^i \bar{z}_\mp^{*j})(1-q^{h_\pm+\frac{1}{2}+m}\bar{q}^{h_\pm+\frac{1}{2}+n}z_\mp^{*i} \bar{z}_\mp^{j})}\ .
\label{eq:scal_pf_shsn}
\end{multline}
Heuristically, one can understand this partition function as follows: the boundary modes of the scalar fields $(\phi_+)_{ij}$ and $(\phi^*_+)_{ij}$
are counted by the factors $q^{h_+ + m}\bar q^{h_+ + n} z_+^i \bar z^{*j}_+$ and, respectively,   $q^{h_+ + m}\bar q^{h_+ + n} z_+^{*i} \bar z^{j}_+$;
the boundary modes of the Dirac fields $(\psi_+)_{ij}$ and $(\psi^*_+)_{ij}$ are counted by the factors  $q^{h_+ +\frac{1}{2}+ m}\bar q^{h_+ + n} z_-^i \bar z^{*j}_+$ and, respectively, $q^{h_+ +\frac{1}{2}+ m}\bar q^{h_+ + n} z_-^{*i} \bar z^{j}_+$; the boundary modes of the Dirac fields $(\psi_-)_{ij}$ and $(\psi^*_-)_{ij}$ are counted by the factors $q^{h_+ + m}\bar q^{h_+ +\frac{1}{2}+ n} z_+^i \bar z^{*j}_-$ and, respectively, $q^{h_+ + m}\bar q^{h_+ +\frac{1}{2}+ n} z_+^{*i} \bar z^{j}_-$ etc.

The matter part~\eqref{matter} of the partition function can be written more compactly with the help of the following  supermatrices of $\mathrm{GL}(\infty|\infty)$
\begin{multline}
U_\pm = \mathrm{diag}(q^{h_\pm} z^1_\pm,\dots,q^{h_\pm} z^{M}_\pm,-q^{h_\pm+\frac{1}{2}} z^1_\mp,\dots,-q^{h_\pm+\frac{1}{2}} z^{M}_\mp,\\
q^{h_\pm+1} z^1_\pm,\dots,q^{h_\pm+1} z^{M}_\pm,-q^{h_\pm+\frac{3}{2}} z^1_\mp,\dots,-q^{h_\pm+\frac{3}{2}} z^{M}_\mp,\dots )\ .
\label{eq:matUpm}
\end{multline}
Similarly, define $U_\pm^*$ by making the replacements $z_\pm^i\mapsto z_\pm^{i*}$ and $\bar{z}_\pm^i\mapsto \bar{z}^{i*}_\pm$ on the r.h.s.\ of eq.~\eqref{eq:matUpm} and, then, $\bar{U}_\pm$ and $\bar{U}^*_\pm$ by putting a bar on everybody.
Here the grading of $\mathrm{GL}(\infty|\infty)$ is induced from the grading of its fundamental representation over $\mathbb{C}^{\mathbb{N}}$ for which the $i$-th component is defined to have the same parity as $[i/M]$, i.e.\ even basis vectors have $U_\pm$ eigenvalues $q^{h_\pm +n}z^i_\pm$, while odd basis vectors have eigenvalues $-q^{h_\pm +n+\frac{1}{2}}z^i_\mp$, where $n$ is a non-negative integer.
With this notation the partition function for the matter fields becomes particularly simple and can be immediately expanded in a sum of $\mathrm{GL}(\infty|\infty)$ Schur functions
with the help of the Cauchy identity
\begin{equation*}
Z_{\text{matter}}^\pm = \frac{1}{\mathrm{sdet}(1-U_\pm\otimes \bar{U}^*_\pm)\,\mathrm{sdet}(1-U^*_\pm\otimes \bar{U}_\pm)} = \sum_{\Lambda,\Xi} s_\Lambda(U_\pm)s_\Lambda(\bar{U}^*_\pm)s_\Xi(U^*_\pm)s_\Xi(\bar{U}_\pm)\ .
\end{equation*}
In this way, we get the following expansion for the partition function~\eqref{eq:totoal_pf}
\begin{equation}
Z^{\text{1-loop}}_{\text{Vasiliev}} = Z_{\text{gauge}} \mathop{\sum_{\Lambda_l,\Lambda_r}}_{\Xi_l,\Xi_r} s_{\Lambda_l}(U_+)s_{\Lambda_l}(\bar{U}^*_+)s_{\Lambda_r}(U^*_+)s_{\Lambda_r}(\bar{U}_+)s_{\Xi_l}(U_-)s_{\Xi_l}(\bar{U}^*_-)s_{\Xi_r}(U_-^*)s_{\Xi_r}(\bar{U}_-)\ ,
\label{eq:total_pf_exp}
\end{equation}
which, as we shall see, is most naturally comparable  with the proposed dual coset theory.



\section{Dual coset theories}

In this section we consider the coset CFT based on the coset algebra
\begin{equation}
\frac{\su(N+M)_k \oplus \so(2NM)_1}{\su(N)_{k+M}\oplus \uu(1)_\kappa}
\label{eq:lcoset}
\end{equation}
and its charge conjugate modular invariant. We compute the partition function and higher spin spectrum in the 't~Hooft limit
\begin{equation}
N,k\to\infty \qquad \text{with} \qquad \lambda = \frac{N}{k+N} \qquad \text{held fixed}
\label{eq:tHooft}
\end{equation}
and find perfect agreement with the corresponding quantities in the extended $\shs[\lambda]$ Vasiliev theory  discussed in sec.~\ref{sec:extended}.
These findings strongly support the holographic duality~\cite{Gaberdiel:2013vva, Creutzig:2013tja} between the classical Vasiliev theory based on $\shs[\lambda]$ and subject to the asymptotic $AdS_3$ boundary conditions~\eqref{eq:bc} on the one hand, and the 't~Hooft limit of the coset CFT \eqref{eq:lcoset} on the other hand.

\subsection{Definition}\label{sec:def_coset}

The coset~\eqref{eq:lcoset} can be obtained from the manifestly $\mathcal{N}=1$ coset
\begin{equation}
\frac{\su(N+M)^{(1)}_{k+N+M}}{\su(N)^{(1)}_{k+N+M}\oplus \mathfrak{u}(1)^{(1)}_\kappa}
\label{eq:n1lcoset}
\end{equation}
 by removing all fermions in the denominator together with the remaining $M^2-1$ free fermions in the numerator. Notice that the last step generally  breaks supersymmetry.
The level $\kappa$ will be specified later.

Let us choose a basis of $\su(N+M)_k$ that respects the decomposition 
\begin{equation}
\su(N+M)_k\simeq \underbrace{\su(N)_k}_{J^A}\oplus\underbrace{\su(M)_k}_{J^I}\oplus \underbrace{\uu(1)}_{J}\oplus \underbrace{(N,\bar M)_{N+M}}_{J^{ai}}\oplus \underbrace{(\bar N ,M)_{-N-M}}_{\bar{J}^{ai}}
\label{eq:dec_sunm}
\end{equation}
where the lower index denotes the $J$-charge. The OPEs of the currents in this basis are given in eq.~\eqref{eq:sumn_OPE}.
The $\so(2NM)_1$ factor in the numerator of eq.~\eqref{eq:lcoset} corresponds to $NM$ Dirac fermions $\psi^{ai}$ and their conjugates $\bar{\psi}^{ai}$, which satisfy the OPEs
\begin{equation}
\psi^{a i}(z) \bar{\psi}^{b j}(w) \sim  \frac{\delta_{a b} \delta_{ij}}{z-w} \sim \bar \psi^{a i}(z)\psi^{b j}(w)\ .
\end{equation}
Out of them one can construct the following currents 
\begin{equation}
K^{A}= t^A_{ab}:\psi^{ai}\bar\psi^{bi}:\ ,\qquad K^{I}= t^{I}_{ij}: \bar \psi^{a i}\psi^{aj}:\ ,\qquad K =\psi^{ai}\bar\psi^{ai}\ ,
\label{eq:twosuM}
\end{equation}
which generate the  current algebra $\su(N)_M\oplus\su(M)_N\oplus \uu(1)_{NM}$. W.r.t.\ that algebra the fermions $\psi^{ai}$ transform in the representation $(N,\bar M)_{1}$, $\bar \psi^{ai}$ in the representation $(\bar N,M)_{-1}$, which  can be seen explicitly from the commutation relations \eqref{eq:sumn_OPE_k}.
The embedding of the denominator into the numerator of \eqref{eq:lcoset} is given by
\begin{equation}
\tilde J^{A}:= J^{A} + K^{A}\ ,\qquad \tilde J:= J + (N+M)K\ ,
\label{eq:embsuN}
\end{equation}
where the coefficient of $K$ in the second equality is determined from the requirement that $J^{ai}$ and $\psi^{ai}$ have the same $\tilde J$-charge --- a property  inherited from the $\mathcal{N}=1$ supersymmetry of the parent theory~\eqref{eq:n1lcoset}.
Our convention for the level of  $\tilde J$ is 
\begin{equation}
\tilde J(z)\tilde J(w)\sim  \frac{\kappa}{(z-w)^2}\ ,\qquad \kappa:= NM(N+M)(k+N+M)\ .
\label{eq:levu1}
\end{equation}
The coset algebra~\eqref{eq:lcoset} is then to be understood as the algebra of normal ordered differential polynomials in the numerator currents that are regular w.r.t.\ the denominator currents.
In particular, the energy-momentum tensor of the coset can be computed by the Goddard-Kent-Olive (GKO) construction \cite{Goddard:1986ee, god} and is given explicitly in eq.~\eqref{eq:cosetT_expl}.

The representations  $(\Lambda;\Xi,l)$ of the coset algebra~\eqref{eq:lcoset}  are defined by the usual GKO construction through the decomposition
\begin{equation}
\Lambda\otimes \mathrm{NS} =\bigoplus_{\Xi,l} (\Lambda;\Xi,l)\otimes \Xi\otimes l\ ,
\label{eq:def_coset_rep}
\end{equation} 
where $\Lambda$ is an integrable weight of $\mathrm{SU}(N+M)_k$ identified with a Young diagram of at most $N+M-1$ rows and $k$ columns,  NS is the Neveu-Schwarz sector for the fermions $\psi^{ai}$ and  $\bar \psi^{ai}$, $\Xi$ is an integrable weight of $\mathrm{SU}(N)_{k+M}$ identified with a Young diagram of at most $N-1$ rows and $k+M$ columns, and $l\in\mathbb{Z}_\kappa$ labels the representation of charge $l$ w.r.t\ the $\uu(1)$-current in eq.~\eqref{eq:embsuN}.
The decomposition~\eqref{eq:def_coset_rep} satisfies the selection rule
\begin{equation}
l \equiv (N+M)|\Xi| - N |\Lambda| \mod N(N+M)\ ,
\label{eq:sel_rule}
\end{equation}
which follows from the requirement that the $\su(N+M)$ weights of the affine highest weight vectors on   both hand sides in eq.~\eqref{eq:def_coset_rep} differ by an element of the $\su(N)$ root lattice plus an element of the $\su(M)$ weight lattice.
There are also  field identifications $(\Lambda;\Xi,l)\simeq (\Lambda';\Xi',l')$, which are explained in \cite{Lerche:1989uy}, but they are irrelevant in the 't~Hooft limit because they  do not give rise to non-trivial identifications.
The characters of the coset representations $(\Lambda;\Xi,l)$ are defined by 
\begin{equation}
b_{\Lambda;\Xi,l}(q,z_+,z_-) := \tr_{(\Lambda;\Xi,l)} q^{L_0} \exp[J^I_0\tr(t^I H_+)+K^I_0\tr(t^I H_-)]\ ,
\label{eq:def_br}
\end{equation}
where $L_0$, $J^I_0$, $K^I_0$ are the zero modes of $T$, $J^I$ and $K^I$, respectively, while
$e^{H_{\pm}}$ are two arbitrary points on the Cartan torus of $\mathrm{SU}(M)$
with eigenvalues $z_\pm\equiv (z^1_\pm,\dots,z^M_\pm)$ in the fundamental representation.
More precisely, if $e_i$ is a basis of the fundamental representation of $\mathrm{SU}(M)$ diagonalizing the Cartan subalgebra and $\epsilon_i$ is the weight of $e_i$ then $z_\pm^i = e^{\epsilon_i(H_\pm)}$.

To complete the definition of the coset CFT we must specify a Hilbert space which glues in a modular invariant way the representations of a left and a right copy of the coset algebra~\eqref{eq:lcoset}.
The simplest choice is given by the charge conjugation modular invariant
\begin{equation}
\mathcal{H}_{\text{coset}}= \bigoplus_{[\Lambda;\Xi,l]}(\Lambda;\Xi,l)\otimes \overline{(\Lambda^*;\Xi^*,l^*)}\ ,
\label{eq:hs_coset}
\end{equation}
where $[\Lambda;\Xi,l]$ denotes the equivalence class of the representation $(\Lambda;\Xi,l)$ under the action of the field identification rules and $(-)^*$ denotes the conjugate representation.
The corresponding partition function is then
\begin{equation}
Z_{\text{coset}} = \sum_{[\Lambda;\Xi,l]}b_{\Lambda;\Xi,l}(q,z_+,z_-)b_{\Lambda^*;\Xi^*,l^*}(\bar q,\bar z_+,\bar z_-)\ .
\label{eq:def_coset_pf}
\end{equation}

\subsection{Partition function}\label{sec:pf_coset}

Let us now compute the 't~Hooft limit of the coset partition function~\eqref{eq:def_coset_pf}.
We  recall that the 't~Hooft limit of the Hilbert space~\eqref{eq:hs_coset} is regularized in such a way that only the representations $\Lambda$ of  $\mathrm{SU}(N+M)$ which appear in a \emph{finite} tensor product of the fundamental representation and its conjugate are taken into account. The same remark applies to $\Xi$. This means that in the limit $\Lambda$ and $\Xi$ can be unambiguously specified by a pair of finite Young diagrams, i.e.\
\begin{equation}
\Lambda\mapsto (\Lambda_l,\Lambda_r)\ ,\qquad  \Xi\mapsto (\Xi_l,\Xi_r)\ ,
\label{eq:doubling}
\end{equation}
where the index $r$ corresponds to the tensor built out of the fundamental representation (covariant) and $l$ to the tensor built out of the conjugate representation (contravariant).
These pairs  determine uniquely the $\uu(1)$-charge
\begin{equation}
l \mapsto (N+M)(|\Xi|_r-|\Xi|_l)-N(|\Lambda_r|-|\Lambda_l|)\ ,
\label{eq:l_charge}
\end{equation}
thus removing in the 't~Hooft limit the mod $N(N+M)$ ambiguity of eq.~\eqref{eq:sel_rule}, see \cite{Candu:2012jq} for a more detailed explanation.
Hence, in the 't~Hooft limit the coset branching functions will  effectively be labelled by the pairs of Young diagrams~\eqref{eq:doubling}
\begin{equation}
b_{\Lambda;\Xi,l} \mapsto b_{(\Lambda_l,\Lambda_r);(\Xi_l,\Xi_r)}
\label{eq:c_br_th}
\end{equation}
and the bulk of this section is dedicated to evaluating their limit.

We shall follow the strategy developed in \cite{Candu:2012jq}, which starts with factoring out the $k$ dependence of the branching functions
\begin{equation}
b_{\Lambda;\Xi,l} \simeq  q^{\frac{1}{2(k+N+M)}\left[\mathrm{Cas}^{N+M}(\Lambda)-\mathrm{Cas}^{N}(\Xi)-\frac{l^2}{NM(N+M)}\right]}a_{\Lambda;\Xi,l} \ ,
\label{eq:red_br}
\end{equation}
where $a_{\Lambda;\Xi,l}$ are free, i.e.\ $k\to\infty$,  branching functions defined by
\begin{multline}
\ch^{N+M}_\Lambda (\imath(z_+,u,v)) \prod_{a=1}^N\prod_{i=1}^{M} \prod_{n=1}^\infty \frac{(1+q^{n-\frac{1}{2}}z^{i*}_- u^a v^{N+M})(1+q^{n-\frac{1}{2}} z^{i}_-u^{a*}v^{*N+M})}
{(1-q^n z_+^{i *} u^a v^{N+M})(1-q^nz_+^iu^{a*} v^{*N+M})}\times\\ \times
\prod_{i,j=1}^{M}\prod_{n=1}^\infty \frac{1}{1-q^n  z_+^{i}z_+^{j*}} =
\sum_{\Xi,l}a_{\Lambda;\Xi,l}(q,z_+,z_-)\times \ch_\Xi^{N}(u)\times \frac{v^l}{\prod_{n=1}^\infty (1-q^n)}\ ,
\label{eq:branching_free}
\end{multline}
see \cite{Candu:2012jq} for more details.
Here $z_+$ is an $\mathrm{SU}(M)$ matrix with eigenvalues $\{z_+^i\}$, $u$ is an $\mathrm{SU}(N)$ matrix with eigenvalues $\{u^a\}$,  $v$ is a $\mathrm{U}(1)$ phase and
$\imath(z_+,u,v)$ denotes the following  embedding of $\SU(M)\times \SU(N)\times \U(1)$ into $\mathrm{SU}(N+M)$
\begin{equation}
\imath(z_+,u,v) = \begin{pmatrix}u v^M & 0\\ 0 & z_+v^{-N}\end{pmatrix}\ .
\label{eq:emb_i}
\end{equation}
We shall now compute in two steps from the basic definition~\eqref{eq:branching_free} the 't~Hooft limit of the free theory branching functions
\begin{equation}
a_{\Lambda;\Xi,l} \mapsto a_{(\Lambda_l,\Lambda_r);(\Xi_l,\Xi_r)}\ ,
\label{eq:a_thooft}
\end{equation}
where the representation labels on both hand sides are related by eqs.~(\ref{eq:doubling}, \ref{eq:l_charge}).

Let us start with  $a_{0;(\Xi_l,\Xi_r)}$. Decomposing every factor separately using the Cauchy identity (see e.g. the appendix of \cite{Candu:2012jq}) we get
\begin{multline}
\prod_{a=1}^N\prod_{i=1}^{M} \prod_{n=1}^\infty \frac{(1+q^{n-\frac{1}{2}}z^{i*}_- \tilde u^a)}
{(1-q^n z_+^{i *} \tilde u^a)}  \times \prod_{a=1}^N\prod_{i=1}^{M} \prod_{n=1}^\infty
\frac{(1+q^{n-\frac{1}{2}} z^{i}_-\tilde u^{a*})}{(1-q^nz_+^i\tilde u^{a*} )}=\\= \sum_{\Lambda_l,\Lambda_r}\ch^N_{(\Lambda_l,0)}(\tilde u)\ch^N_{(0,\Lambda_r)}(\tilde u)\ s_{\Lambda_l}(U_{\frac{1}{2}})s_{\Lambda_r}(U^*_{\frac{1}{2}})\ ,
\label{eq:dec_step1}
\end{multline}
where $\{\tilde u^a\}$ are the eigenvalues of the $\U(N)$ matrix $\tilde u = u v^{N+M}$, the sum runs over all Young diagrams $\Lambda_l$, $\Lambda_r$ of at most $N$ rows,  $\ch^N_{(\Lambda_l,0)}$ is the character of the irreducible purely contravariant $\U(N)$ tensor of shape $\Lambda_l$,
$\ch^N_{(0,\Lambda_r)}$  is the character of the irreducible purely covariant $\U(N)$ tensor of shape $\Lambda_r$, $s_\Lambda$ are $\mathrm{GL}(\infty|\infty)$ Schur functions, 
and $U_{\frac{1}{2}}$ is a $\mathrm{GL}(\infty|\infty)$ matrix 
\begin{multline}
U_{\frac{1}{2}} = \mathrm{diag}(-q^{\frac{1}{2}} z^1_-,\dots,-q^{\frac{1}{2}} z^{M}_-,
q z^1_+,\dots,q z^{M}_+,-q^{\frac{3}{2}} z^1_-,\dots,-q^{\frac{3}{2}} z^{M}_-, q^2 z^1_+,\dots,q^2 z^{M}_+,\dots)\ ,
\label{eq:matU12}
\end{multline}
which is obtained from $U_+$ in eq.~\eqref{eq:matUpm} by setting $h_+=0$ and removing the first $M$ rows and columns while keeping the parity of the remaining entries unchanged.
If we  now use the explicit form of the $\mathrm{U}(N)$ Clebsch-Gordan coefficients \cite{King:1971rs}
\begin{equation}
c^{(\Xi_l,\Xi_r)}_{(\Lambda_l,0)(0,\Lambda_r)} = \sum_\Pi c^{\Lambda_l}_{\Xi_l \Pi}c^{\Lambda_r}_{\Xi_r \Pi}\ ,
\label{eq:CG_tf}
\end{equation}
where $c^\Lambda_{\Xi\Pi}$ are the Littlewood-Richardson coefficients, and the basic property  $s_\Xi s_\Pi = \sum_\Lambda c^{\Lambda}_{\Xi\Pi} s_\Lambda$ of the Schur functions, then
the r.h.s.\ of eq.~\eqref{eq:dec_step1} becomes
\begin{equation}
\sum_{\Xi_l,\Xi_r,\Pi} \ch^N_{(\Xi_l,\Xi_r)}(\tilde u)s_{\Xi_l}(U_{\frac{1}{2}})s_{\Pi}(U_{\frac{1}{2}})s_{\Xi_r}(U^*_{\frac{1}{2}})s_{\Pi}(U^*_{\frac{1}{2}}) = \sum_{\Xi_l,\Xi_r}\frac{\ch^N_{(\Xi_l,\Xi_r)}(\tilde u)s_{\Xi_l}(U_{\frac{1}{2}})s_{\Xi_r}(U^*_{\frac{1}{2}})}{\mathrm{sdet}(1-U_{\frac{1}{2}}\otimes U_{\frac{1}{2}}^*)}\ .
\label{eq:pf_step2}
\end{equation}
In this  equality  we have used the Cauchy identity to evaluate the sum over $\Pi$.
Next, taking into account the  character relation
\begin{equation}
\ch^N_{(\Xi_l,\Xi_r)}(\tilde u) = \ch^N_{(\Xi_l,\Xi_r)}(u)\times  v^{(N+M)(|\Xi_r|-|\Xi_l|)}
\label{eq:u1_fact}
\end{equation}
 and eq.~\eqref{eq:l_charge} we can compare the  left hand sides of  eqs.~(\ref{eq:pf_step2},  \ref{eq:branching_free}) 
to obtain
\begin{multline}
a_{0;(\Xi_l,\Xi_r)}(q,z_+,z_-) = s_{\Xi_l}(U_{\frac{1}{2}})s_{\Xi_r}(U^*_{\frac{1}{2}})\times
\prod_{n=1}^\infty \frac{1-q^n}{\prod_{i,j=1}^{M}(1-q^n  z_+^{i}z_+^{j*})}\times \\ \times
\prod_{s=1}^\infty \prod_{n=s}^\infty \prod_{i,j=1}^{M} 
\frac{(1+q^{n+\frac{1}{2}}z^i_{+} z_{-}^{j*})(1+q^{n+\frac{1}{2}}z^{i*}_{+} z_{-}^{j})}{(1-q^{n+1} z_+^i z_+^{j*})(1-q^n z_-^i {z_-^{j*}})}\ ,
\label{eq:a0xi}
\end{multline}
where the first product is inherited from the $u$-independent products in eq.~\eqref{eq:branching_free}, while the second product comes from the superdeterminant in eq.~\eqref{eq:pf_step2}.
In particular, notice that the vacuum character of the theory is given by
\begin{equation}
b_{0;0}= a_{0;0} = \prod_{n=1}^\infty(1-q^n)\times \prod_{s=1}^\infty \prod_{n=s}^\infty \prod_{i,j=1}^{M} 
\frac{(1+q^{n+\frac{1}{2}}z^i_{+} z_{-}^{j*})(1+q^{n+\frac{1}{2}}z^{i*}_{+} z_{-}^{j})}{(1-q^{n} z_+^i z_+^{j*})(1-q^n z_-^i {z_-^{j*}})}\ .
\label{eq:vac_lcoset}
\end{equation}

In a second step, consider the general branching functions $a_{(\Lambda_l,\Lambda_r);(\Xi_l,\Xi_r)}$.
From the definition~\eqref{eq:branching_free} we get
\begin{equation}
a_{(\Lambda_l,\Lambda_r);(\Xi_l,\Xi_r)}(q,z_+,z_-) = \mathop{\sum_{\Phi_l,\Phi_r,\Psi_l}}_{\Psi_r,\Pi_l,\Pi_r} r^{(\Lambda_l,\Lambda_r)}_{(\Phi_l,\Phi_r)(\Psi_l,\Psi_r)}c^{(\Pi_r,\Pi_l)}_{(\Psi_l,\Psi_r)(\Xi_r,\Xi_l)}\ch^{M}_{(\Phi_l,\Phi_r)}(z_+)a_{0;(\Pi_l,\Pi_r)}(q,z_+,z_-)\ .
\label{eq:acomplete}
\end{equation}
To obtain this relation we have used the $\U(N+M)\downarrow\U(N)\times \U(M)$ restriction rules (cf.\ \ref{eq:emb_i})
\begin{equation}
\ch^{N+M}_{(\Lambda_l,\Lambda_r)}(\imath(z_+,u,v)) = \sum_{\Phi_l,\Phi_r,\Psi_l,\Psi_r}r^{(\Lambda_l,\Lambda_r)}_{(\Phi_l,\Phi_r)(\Psi_l,\Psi_r)}\ch^{M}_{(\Phi_l,\Phi_r)}(z_+ v^{-N}) \ch^N_{(\Psi_l,\Psi_r)}(u v^M)
\label{eq:restr_def}
\end{equation}
where the explicit form of the restriction coefficients was given  in  \cite{King:1975vf}
\begin{equation}
r^{(\Lambda_l,\Lambda_r)}_{(\Phi_l,\Phi_r)(\Psi_l,\Psi_r)}=\sum_{\sigma,\tau,\rho}c^{\Lambda_l}_{\Phi_l\sigma}c^{\Lambda_r}_{\Phi_r\tau}c^{\sigma}_{\Psi_l\rho}c^{\tau}_{\Psi_r\rho}
\label{eq:restr}
\end{equation}
repeatedly used eq.~\eqref{eq:u1_fact}, a relation following from eq.~\eqref{eq:restr}
\begin{equation}
|\Lambda_l|-|\Phi_l|-|\Psi_l| = |\Lambda_r|-|\Phi_r|-|\Psi_r|\ ,
\label{eq:disc_symm}
\end{equation}
and the obvious symmetry of the $\mathrm{U}(N)$ Clebsch-Gordan coefficients, which can be found in~\cite{King:1971rs} 
\begin{equation}
c^{(\Xi_l,\Xi_r)}_{(\Psi_l,\Psi_r)(\Pi_l,\Pi_r)}=c^{(\Pi_r,\Pi_l)}_{(\Psi_l,\Psi_r)(\Xi_r,\Xi_l)} = \sum_{\pi,\rho,\sigma,\tau} c^{(\pi,\rho)}_{(\Psi_l,0)(0,\Pi_r)}c^{(\sigma,\tau)}_{(\Pi_l,0)(0,\Psi_r)}c^{\Xi_l}_{\pi\sigma}c_{\rho\tau}^{\Xi_r}\ .
\label{eq:symmetry_CG}
\end{equation}

To make good use of the formula~\eqref{eq:acomplete} we now need to take into account the emergence  of null vectors in the 't~Hooft limit. 
By analogy with \cite{Gaberdiel:2011zw}, we shall make the usual assumption that their removal 
is equivalent to declaring that in the limit, which assumes that $N\to \infty$, the fundamental representations of $\SU(N+M)$ and $\SU(N)$ do not talk to their duals.
This prescription is equivalent to the following factorization of the restriction and Clebsch-Gordan coefficients in eq.~\eqref{eq:acomplete} (cf. \ref{eq:restr}, \ref{eq:symmetry_CG})
\begin{align}\notag
r^{(\Lambda_l,\Lambda_r)}_{(\Phi_l,\Phi_r)(\Psi_l,\Psi_r)}&\mapsto c^{\Lambda_l}_{\Phi_l\Psi_l} c^{\Lambda_r}_{\Phi_r\Psi_r}\ ,\\
c^{(\Pi_r,\Pi_l)}_{(\Psi_l,\Psi_r)(\Xi_r,\Xi_l)}&\mapsto c^{\Pi_r}_{\Psi_l\Xi_r}c^{\Pi_l}_{\Psi_r\Xi_l}\ ,
\label{eq:null_vecs_rem}
\end{align}
where we have used that the $\mathrm{U}(N)$ Clebsch-Gordan and restriction coefficients of purely covariant or purely contravariant tensors coincide with the Littlewood-Richardson coefficients.
Making these replacements in eq.~\eqref{eq:acomplete} we can evaluate all sums to
\begin{equation}
a_{(\Lambda_l,\Lambda_r);(\Xi_l,\Xi_r)} = s_{\Lambda_l}(U^*_0)s_{\Lambda_r}(U_0)s_{\Xi_l}(U_{\frac{1}{2}})s_{\Xi_r}(U^*_{\frac{1}{2}})\, a_{0;0}\ ,
\label{eq:result_agen}
\end{equation}
where $U_0$ is the matrix $U_\pm$ with $h_\pm=0$ and
we have used the fact that the $\mathrm{GL}(\infty+M|\infty)\downarrow \mathrm{GL}(M)\times \mathrm{GL}(\infty|\infty)$ restriction coefficients of purely covariant tensors coincide with the Littlewood-Richardson coefficients  (see e.g. \cite{Candu:2012jq})
\begin{equation}
s_{\Lambda}(U_0) = \sum_{\Phi,\Psi}c^{\Lambda}_{\Phi\Psi} \ch^{M}_{\Phi}(z_+)s_{\Psi}(U_\frac{1}{2})\ .
\label{eq:summ_restr}
\end{equation}

We are now ready to evaluate the r.h.s.\ of eq.~\eqref{eq:red_br}.
Approximating the $\SU(N+M)$ and $\SU(N)$ Casimirs with their dominant terms (see e.g.\  \cite{Candu:2012jq})
\begin{equation}
\frac{\mathrm{Cas}^{N+M}(\Lambda)-\mathrm{Cas}^{N}(\Xi)}{2(k+N+M)}-\frac{l^2}{2(k+N+M)NM(N+M)} \simeq \frac{\lambda}{2} (|\Lambda_l|+|\Lambda_r|-|\Xi_l|-|\Xi_r|)\ ,
\label{eq:cas_assympt}
\end{equation}
one can absorb the overall power of $q$ in eq.~\eqref{eq:red_br} in the entries of the matrices $U_0$, $U_{\frac{1}{2}}$, i.e.
\begin{equation}
s_\Lambda (U_+) = q^{\frac{\lambda}{2}|\Lambda|}s_\Lambda(U_0)\ ,\qquad s_{\Xi^t} (U_-) = q^{-\frac{\lambda}{2}|\Xi|}s_\Xi(U_\frac{1}{2})\ ,
\label{eq:snum}
\end{equation}
where in the second case the transpose comes from the opposite grading of the otherwise equal matrices $U_-$ and $q^{-\frac{\lambda}{2}}U_{\frac{1}{2}}$, see the appendix of \cite{Candu:2012jq} for more details.
Thus, putting everything together we get
\begin{equation}
b_{(\Lambda_l,\Lambda_r);(\Xi_l,\Xi_r)} = s_{\Lambda_l}(U^*_+)s_{\Lambda_r}(U_+)s_{\Xi^t_l}(U_{-})s_{\Xi^t_r}(U^*_{-})\, a_{0;0}\ .
\label{eq:branching_complete}
\end{equation}
It is then obvious that the coset partition function~\eqref{eq:def_coset_pf} regularized as
\begin{equation}
Z^{\text{'t~Hooft}}_{\text{coset}} :=
\sum_{\Lambda_l,\Lambda_r,\Xi_l,\Xi_r} b_{(\Lambda_l,\Lambda_r);(\Xi_l,\Xi_r)}(q,z_+,z_-)  b_{(\Lambda_r,\Lambda_l);(\Xi_r,\Xi_l)}(\bar q,\bar z_+, \bar z_-)
\label{eq:def_coset_pf_lim}
\end{equation}
gives exactly the higher spin partition function~\eqref{eq:total_pf_exp} 
\begin{equation}
Z^{\text{'t~Hooft}}_{\text{coset}} = Z^{\text{1-loop}}_{\text{Vasiliev}}\ .
\label{eq:pf_comparison}
\end{equation}

\subsection{Higher spin spectrum}\label{sec:hs_spec_c1}

The generating function for the spectrum of the $\mathcal{W}$-algebra of a coset is in general given by the vacuum character, see \cite{Bouwknegt:1992wg}. For the coset~\eqref{eq:lcoset}, the 't~Hooft limit of the vacuum character   was computed in eq.~\eqref{eq:vac_lcoset}.
From its simple product form one can immediately conclude that the $\mathcal{W}$-algebra of the coset is freely generated in the 't~Hooft limit. Taking into account that a spin $s$
current which is not subject to any constraints contributes to the vacuum character with a factor
\begin{equation*}
\prod_{n=s}^\infty \frac{1}{1-q^n}\ ,
\end{equation*}
if it is bosonic, i.e.\ $s\in\mathbb{N}$, or with a factor
\begin{equation*}
\prod_{n=s-\frac{1}{2}}^\infty (1+q^{n+\frac{1}{2}})\ ,
\end{equation*}
if it is fermionic, i.e.\ $s\in\mathbb{N}-\frac{1}{2}$, one can clearly read off from ~\eqref{eq:vac_lcoset} the following spectrum of generators (cf. \ref{eq:dec_shsn})
\begin{align}\label{eq:coset_hsc}
s=1:&\quad  (adj, 0) \oplus(0,adj)\oplus (0,0)\\ \notag
s\in \mathbb{N}+\tfrac{1}{2}:&\quad (f,f^*)\oplus(f^*,f)\\ \notag
s\in\mathbb{N}+1:&\quad (adj,0)\oplus(0,adj)\oplus 2 (0,0)\ ,
\end{align}
where the the pair of labels in the brackets denotes the transformation properties of these generators w.r.t.\ the two $\su(M)$'s generated by $J^I_0$ and $K^I_0$, respectively.
In total we have $2\,M^2-1$ generators with spin $1$ and $2\,M^2$ generators for every half-integer spin $s \geq 3/2$.
This spin content matches precisely the spectrum of the asymptotic symmetry algebra of the higher spin theory~\eqref{eq:gfix}.

Let us now construct the currents in eq.~\eqref{eq:coset_hsc} explicitly. Clearly, the spin 1 currents always exist even  at finite central charge and can be identified with
\begin{equation}
J^I\ ,\qquad  K^I\ ,\qquad  U= \frac{J - k K}{k+N+M}\ ,
\label{eq:coset_s1c}
\end{equation}
respectively. Similarly, the spin $s=3/2$ currents are given by
\begin{equation}
  W^{(3/2)-}_{ij}=\psi^{ai} \bar{J}^{aj}\ ,\qquad  W^{(3/2)+}_{ij}=J^{a i} \bar{\psi}^{a j}\ ,
\label{eq:spin32_coset}
\end{equation}
respectively, where $i,j$ are free indices. They are both Virasoro primary and affine primary w.r.t.\ the currents $J^I$ and $K^I$.
The problem of constructing  the currents of spin $s\geq 2$ explicitly at finite central charge is still feasible, but quickly becomes technically very complicated with increasing spin, see \cite{Ahn:2013bhw} for an illustrative example. Let us explain the main idea behind this construction.

The first step is to  introduce the covariant derivative, which sends affine primaries of the denominator subgroup into affine primaries \cite{deBoer:1993gd}.
Thus, given a set of $\su(N)_{k+M}\oplus\uu(1)_\kappa$ affine primary fields $\Phi^a$ transforming in the representation
$\rho^A_{ab}\equiv \rho(t^A)_{ab}$ of $\su(N)$ and of $\tilde J$-charge $Q$, i.e.\
\begin{equation}
\tilde J^A(z)\Phi^a(w) \sim \frac{\rho^A_{ba}\Phi^b(w)}{z-w}\ ,\quad \tilde J(z)\Phi^a(w)\sim \frac{Q\Phi^a(w)}{z-w}\ ,
\label{eq:aff_prim_def}
\end{equation}
their covariant derivative is defined as follows
\begin{equation}
\mathcal{D}\Phi^a = \partial \Phi^a -\frac{1}{k+N+M}\left[ \frac{Q(\tilde J \Phi^a)}{NM(N+M)}+\rho^A_{ba}(\tilde J^A\Phi^b)\right]\ .
\label{eq:cov_der_def}
\end{equation}
Their OPEs with the  $\su(N)_{k+M}\oplus\uu(1)_\kappa$ currents are again given by eq.~\eqref{eq:aff_prim_def}, except $\Phi^a$ gets replaced by $\mathcal{D}\Phi^a$.
Now we can write the coset currents of spin $s\geq 2$ as follows
\begin{align} \label{suSpins}
s\in \mathbb{N}+\tfrac{3}{2}:&&  W^{(s)-}_{ij}&= ( \psi^{a i}\mathcal{D}^{s-\frac{3}{2}} \bar{J}^{a j})+\cdots  \ ,&   W^{(s)+}_{ij}&=(J^{a i} \mathcal{D}^{s-\frac{3}{2}} \bar{\psi}^{a j}) +\cdots\ ,\\ \notag
s\in\mathbb{N}+1:&&   W^{(s)-}_{ij}&=(\psi^{a i} \mathcal{D}^{s-1}  \bar{\psi}^{a j})+\cdots \ ,&   W^{(s)+}_{ij}&=  (J^{a i} \mathcal{D}^{s-2}  \bar{J}^{a j})+\cdots \ ,
\end{align}
where the dots contain ``lower order'' terms in the following sense.
The contraction of $\su(N)$ indices of the dominant terms  ensures that no first order poles can appear in their OPEs with the denominator currents~\eqref{eq:embsuN}.
If the dominant terms do not require a normal ordering then all higher order poles will also vanish, but this is the case only for the spin $s=3/2$ currents~\eqref{eq:spin32_coset}.
In all other cases the higher order poles will not vanish and in order to remove these poles one must correct the dominant terms by descendants of the operators appearing in the singular part of their OPE.
This is precisely what makes the explicit construction of higher spin currents technically complicated.

However, in the limit $k\to\infty$, which is defined in such a way that only the zero modes of the currents $J^I$, $J^A$ and $J$ survive, things simplify considerably: the currents $J^{ai}/\sqrt{k}$, $\bar J^{ai}/\sqrt{k}$ become abelian, the covariant derivative~\eqref{eq:cov_der_def} simplifies to the usual derivative, the normal ordering in eq.~\eqref{suSpins} becomes a Wick normal ordering and all terms hidden by the dots vanish.



\section{Kazama-Suzuki models}

In this section we explain how the refining of the asymptotic boundary conditions~\eqref{eq:bc} of the $\shs_M[\lambda]$ Vasiliev theory of sec.~\ref{sec:extended} can change the CFT dual from the coset~\eqref{eq:lcoset} to the  $\mathcal{N}=2$ Kazama-Suzuki coset
\begin{equation}
\frac{\su(N+M)_{k}\oplus\so(2NM)_1}{\su(N)_{k+M}\oplus \su(M)_{k+N}\oplus\mathfrak{u}(1)}
\label{eq:ks}
\end{equation}
 of complex Grassmannian type. The parameter $\lambda$ is defined as before, see eq.~\eqref{eq:tHooft}.
The new boundary  conditions of the Vasiliev theory do not change the asymptotic AdS geometry   because they differ by $\mathcal{O}(\rho^0)$ terms from the standard AdS boundary conditions~\eqref{eq:bc}.
Their only effect is to constrain an $\su(M)$ subset of the vector fields of the theory and, on the coset side, this corresponds to gauging  the parent coset theory~\eqref{eq:lcoset} by an additional  $\su(M)_{N+k}$ factor.

The section is structured as follows.  First, we present the Kazama-Suzuki model as a gauged version of the previous coset \eqref{eq:lcoset} and then reformulate it as a constrained system.
Next, we discuss the $\mathcal{W}$-algebra of the Kazama-Suzuki coset and emphasize some of its new features. Finally, we interpret the constraints as being part of the asymptotic boundary conditions of the dual $\shs_M[\lambda]$ Vasiliev theory.

\subsection{Partition function}\label{sec:def_KS}

The coset algebra~\eqref{eq:lcoset} contains two current subalgebras: $\su(M)_k$ generated by the currents $J^I$ and $\su(M)_N$ generated by the currents $K^I$,
where we have used the notation of sec.~\ref{sec:def_coset}.
If we now gauge the coset~\eqref{eq:lcoset} by the diagonally embedded current algebra $\su(M)_{k+N}$, generated by the currents
\begin{equation}
\tilde J^I = J^I + K^I\ ,
\label{eq:emb_sum}
\end{equation}
then we obtain the Kazama-Suzuki coset~\eqref{eq:ks}.
This coset can also be obtained directly from the  manifestly $\mathcal{N}=1$ complex Grassmannian coset
\begin{equation}
\frac{\su(N+M)^{(1)}_{k+N+M}}{\su(N)^{(1)}_{k+N+M}\oplus \su(M)^{(1)}_{k+N+M}\oplus\mathfrak{u}(1)^{(1)}}\ ,
\label{eq:ksn1}
\end{equation}
by removing all fermions in the denominator and according to Kazama and Suzuki \cite{Kazama:1988qp} it has an increased $\mathcal{N}=2$ supersymmetry.

The representations $(\Lambda;\Pi,\Xi,l)$ of the coset algebra~\eqref{eq:ks} in the Neveu-Schwarz sector are realized on the multiplicity spaces of the $\su(M)_{k+N}$  representations realized inside   the 
representations  $(\Lambda;\Xi,l)$ of the coset~\eqref{eq:lcoset}
\begin{equation}
 (\Lambda;\Xi,l) =\bigoplus_\Pi (\Lambda;\Pi,\Xi,l)\otimes \Pi\ ,
\label{eq:ks_rep}
\end{equation}
where $\Pi$ is an integrable weight of $\su(M)_{k+N}$ identified with a Young diagram of at most $M-1$ rows and $k+N$ columns which must satisfy a selection rule similar to eq.~\eqref{eq:sel_rule}, see~\cite{Lerche:1989uy}.
The branching functions of the Kazama-Suzuki coset are defined as
\begin{equation}
b_{\Lambda;\Pi,\Xi,l}(q) := \tr_{(\Lambda;\Pi,\Xi,l)} q^{L^{\KS}_0} \ ,
\label{eq:def_br_KS}
\end{equation}
where $L_0^{\KS}$ is the zero mode of the energy momentum tensor of the Kazama-Suzuki coset obtained from eq.~\eqref{eq:cosetT_expl} by subtracting the Sugawara energy momentum tensor of $\su(M)_{k+N}$.
Using again the charge conjugate modular invariant of the coset algebra~\eqref{eq:ks} to define the Hilbert space
\begin{equation}
\mathcal{H}_{\KS}= \bigoplus_{[\Lambda;\Pi,\Xi,l]}(\Lambda;\Pi,\Xi,l)\otimes \overline{(\Lambda^*;\Pi^*,\Xi^*,l^*)}\ ,
\label{eq:hs_coset_KS}
\end{equation}
where $[\Lambda;\Pi,\Xi,l]$ denotes the equivalence class of the representation $(\Lambda;\Pi,\Xi,l)$ under the action of the field identification rules \cite{Lerche:1989uy},
the corresponding partition function becomes
\begin{equation}
Z_{\KS} = \sum_{[\Lambda;\Pi,\Xi,l]}b_{\Lambda;\Pi,\Xi,l}(q)b_{\Lambda^*;\Pi^*,\Xi^*,l^*}(\bar q)\ .
\label{eq:def_coset_pf_KS}
\end{equation}
As before,  we ignore the field identification rules because in the 't~Hooft limit they  do not give rise to non-trivial identifications.

Let us now look at the 't~Hooft limit of the Kazama-Suzuki coset.
The Hilbert space of the theory is regularized in the same way as in sec.~\ref{sec:pf_coset}, i.e.\ the triplet of labels $(\Lambda,\Xi,l)$ is replaced by the pair of finite Young diagrams~\eqref{eq:doubling}, while the range of the Young diagram $\Pi$ extends to all finite  partitions of at most $M-1$ rows.
Taking the 't~Hooft limit of the character of both hand sides in eq.~\eqref{eq:ks_rep} with the techniques of \cite{Candu:2012jq} we obtain the relation
\begin{equation}
b_{\Lambda;\Xi,l}(q,z_+,z_-)\Big\vert_{z_\pm=z} = \sum_{\Pi}b_{\Lambda;\Pi,\Xi,l}(q)\times \frac{\ch^M_\Pi(z)}{\prod_{n=1}^\infty(1-q^n)^{-1}\prod_{i,j=1}^M(1-q^n z^i z^{*j})}\ ,
\label{eq:basic_coset_KS_relation}
\end{equation}
where the branching function on the l.h.s.\ is given by eqs.~(\ref{eq:c_br_th}, \ref{eq:branching_complete}) and we have identified the chemical potential $z_+$ associated to $\su(M)_k$ with the chemical potential $z_-$ associated to  $\su(M)_N$ because the $\su(M)_{k+N}$ factor in the denominator of the Kazama-Suzuki coset is embedded diagonally into $\su(M)_k\oplus\su(M)_N$, see eq.~\eqref{eq:emb_sum}.
Taking a look at the partition functions~(\ref{eq:def_coset_pf}, \ref{eq:def_coset_pf_KS}) and then at eq.~\eqref{eq:basic_coset_KS_relation} and its right moving counterpart we conclude that in the 't~Hooft limit the following relation holds
\begin{equation}
Z^{\text{'t~Hooft}}_{\KS} = \left[Z^{\text{'t~Hooft}}_{\text{coset}}\Big\vert_{z_\pm=\bar z_\pm=z}\times \bigg|\prod_{n=1}^\infty(1-q^n)^{-1}\prod_{i,j=1}^M(1-q^n z^i z^{*j})\bigg|^2\right]^{\su(M)\text{--invariant}}\ ,
\label{eq:relation_pf}
\end{equation}
where the invariance condition is imposed by expanding the term in the brackets in terms of $\su(M)$ characters $\ch^M_\Pi(z)$ and then restricting to the $\Pi=0$ piece.
Put differently, the partition function of the Kazama-Suzuki coset can be obtained from the partition function of the previously considered coset~\eqref{eq:lcoset} by removing the
contribution of the $\su(M)_{k+N}$ currents $J^I+K^I$ and of their right moving counterparts $\bar J^I+\bar K^I$, and then by imposing the singlet condition w.r.t.\ the global $\su(M)$
generated by their zero modes $J^I_0+K^I_0+\bar J^I_0 + \bar K^I_0$.

On the cylinder, the above relation between the two coset theories, at least in the 't~Hooft limit, can be reformulated in the following way: the Kazama-Suzuki coset is equivalent to the  parent coset theory~\eqref{eq:lcoset} subject to the constraints
\begin{equation}
J^I+K^I+\bar J^I + \bar K^I=0\ .
\label{eq:constr_KS}
\end{equation}
In order to prove this, we shall use the fact that the 't~Hooft limit can be interpreted as a classical limit, see sec.~\ref{sec:hsKS} for more details.
Thus, in the 't~Hooft limit the  $\su(M)$ currents $ J^I$, $ K^I$, $ \bar J^I$, $ \bar K^I$ 
become classical fields  and
 we can treat the constraints~\eqref{eq:constr_KS} by standard classical methods. If we develop eq.~\eqref{eq:constr_KS} in Fourier modes on the cylinder, then we get
\begin{equation}
J^I_0+K^I_0+\bar J^I_0 + \bar K^I_0=0\ ,\quad
 J^I_m + K^I_m = 0 \ , \quad \bar J^I_m + \bar K^I_m = 0\ ,\qquad m\neq 0\ .
\label{eq:mode_constr}
\end{equation}
One can easily check that the first type of constraints are first class, while the last two types of constraints are second class.
The second class constraints can be dealt with simply by restricting the phase space of the unconstrained theory to the constraint surface and then replacing the Poisson bracket by the Dirac bracket.
This procedure produces the second factor in the product enclosed in  brackets in eq.~\eqref{eq:relation_pf}.
The first class constraints, on the other hand, after restricting to the constraint surface leave behind residual gauge transformation which are
generated by the constraints themselves. To deal with them one must restrict to gauge invariant quantities, i.e.\ $\su(M)$ invariants  in this case.
This reproduces the $\su(M)$ singlet condition in eq.~\eqref{eq:relation_pf}.

\subsection{Higher spin spectrum}\label{sec:hsKS}


In this section we shall describe the $\mathcal{W}$-algebra of the Kazama-Suzuki coset which, as it turns out, differs considerably from the previously considered algebras \cite{Gaberdiel:2011wb,  Gaberdiel:2012ku, 
Candu:2012tr, Hanaki:2012yf,  Candu:2012ne,  Beccaria:2013wqa, Candu:2013uya}.
We shall work in the  't~Hooft limit which is  interpreted here as a classical limit after identifying  $\hbar\propto 1/N$, see \cite{Ahn:2012fz, Candu:2013uya}.
Thus, after rescaling the generators of sec.~\ref{sec:hs_spec_c1} by an appropriate power of $\hbar$ and taking the 't~Hooft limit they become classical fields that commute with each other.\footnote{As usual, the order $\mathcal{O}(\hbar)$ term in the quantum commutator defines the Poisson bracket.}
We shall denote them in the same way as before.

In the first approach,  the higher spin fields of the Kazama-Suzuki coset are defined as polynomials in the generators of the parent coset theory~\eqref{eq:lcoset}
\begin{equation}
J^I\ ,\quad K^I\ ,\quad  U\ ,\quad \text{ and }\quad  W^{(3/2)\pm}_{ij}\ , \quad W^{(2)\pm}_{ij}\ ,\quad W^{(5/2)\pm}_{ij}\ , \quad W^{(3)\pm}_{ij}\ ,\dots\ ,
\end{equation}
which are regular, i.e.\ Poisson-commute, with the currents $\tilde J^I$.
In order to build such polynomials  one can assemble the $s=1$  affine primaries w.r.t.\ the currents $\tilde J^I$ into a multiplet transforming like the tensor product $f\otimes f^*$
\begin{equation}
W^{(1)\pm}_{ij}\equiv W^{(1)}_{ij} = t^{I}_{ji}\left(\frac{J^I}{k}-\frac{K^I}{N}\right) + \delta_{ij}\frac{U}{k+N}\ ,
\end{equation}
where to normalize every term we have used the fact that the currents $J^I/k$, $K^I/N$ and $U/(k+N)$ have a well defined classical limit.
Furthermore, we shall assume that one can  redefine all other higher spin fields $W^{(s)\pm}_{ij}$ with $s\geq 3/2$ so that they  become $\tilde J^I$-affine primaries transforming in the the $f\otimes  f^*$
of $\su(M)$. Then, the covariant derivatives
\begin{equation}
\D W^{s\pm}_{ij} = \partial W^{s\pm}_{ij} - \frac{\tilde J^I}{k+N}(t^I_{li}W^{(s)\pm}_{lj}- t^I_{jl}W^{(s)\pm}_{il})\ ,
\label{eq:covKS}
\end{equation}
and their higher order powers are also $f\otimes  f^*$ affine primary. Now, because we are in a classical setting, it is very easy to impose the regularity condition w.r.t.\ $\tilde J^I$. Defining the $M\times M$ matrix  $\hat W^{(s)\varepsilon}$ with matrix elements $W^{(s)\varepsilon}_{ij}$, one can construct a manifestly $\su(M)$ invariant field simply by taking traces
\begin{equation}
\tr \D^{n_1} \hat W^{(s_1)\varepsilon_1} \D^{n_1} \hat W^{(s_2)\varepsilon_2} \cdots \D^{n_L} \hat W^{(s_L)\varepsilon_L}\ .
\label{eq:singletr}
\end{equation}
where $s_i\geq 1$, $\varepsilon_i=\pm 1$.
Now because the Poisson bracket satisfies the Leibniz rule, these $\su(M)$ invariant fields will also Poisson-commute with all the modes of $\tilde J^I$ and not only with the zero modes.
Thus, they are  higher spin fields of the Kazama-Suzuki coset.
The first fundamental theorem of classical invariant theory \cite{weyl} then insures that the single trace fields~(\ref{eq:singletr}) generate the entire $\mathcal{W}$-algebra of the  Kazama-Suzuki coset in the 't~Hooft limit.

In the second approach the $\mathcal{W}$-algebra of the Kazama-Suzuki coset is defined as the $\mathcal{W}$-algebra of the parent theory subject to the constraints
\begin{equation}
\tilde J^I=0\ .
\label{eq:w_alg_constr}
\end{equation}
From the Poisson brackets
\begin{equation}\label{eq:diracC}
 C^{IK}_{mn} :=\{\tilde J^I_m, \tilde J^{K}_n\}\big\vert_{\tilde J^{L}=0}= (k+N)m \delta_{m+n,0}\delta^{IK}
\end{equation}
where $\tilde J^I_m$ are the modes of $\tilde J^I$, one clearly sees that the constraints $\tilde J^I_0= 0$ are first class, while the constraints $\tilde J^I_m=0$ with $m\neq 0$ are second class.
To resolve the  second class constraints one must restrict to the constraint surface and then replaces the Poisson bracket inherited from the coset theory with the Dirac bracket
\begin{equation}
\{F,G\}_* = \{F,G\} -\sum_{m,n\neq 0} \{F,\tilde J^I_m\} (C^{-1}_2)^{IK}_{mn}\{\tilde J^{K}_{n}, G\} 
\label{eq:DB}
\end{equation}
where $F$ and $G$ are functionals on the phase space of the parent theory and $C_2$ is the restriction of the matrix~\eqref{eq:diracC} to the second class constraints.
To solve the first class constraints, in addition to imposing $\tilde J^I=0$, one must
restrict to observables Poisson commuting with $\tilde J^I_0$.
Thus, the higher spin fields of the Kazama-Suzuki coset generating the $\mathcal{W}$-algebra will again be given by eq.~\eqref{eq:singletr}, but now the covariant derivatives must be replaced by simple derivatives and
the Poisson bracket inherited from the parent theory by the Dirac bracket~\eqref{eq:DB}.

The  equivalence between these two apparently different presentations of the $\mathcal{W}$-algebra of the Kazama-Suzuki coset can be seen as follows.
Let us identify the generators of the second approach with the restriction of the generators~\eqref{eq:singletr}  to the constraint surface~\eqref{eq:w_alg_constr}. Then the second term in eq.~\eqref{eq:DB} vanishes and the Poisson brackets in both approaches manifestly agree with each other.

In conclusion, let us notice that when $M>1$ the generators~\eqref{eq:singletr} of the $\mathcal{W}$-algebra of the Kazama-Suzuki coset cannot be free because they must satisfy the (infinitely many) relations dictated by the second fundamental theorem of classical invariant theory \cite{weyl}.
For this reason, it is clear that the $\mathcal{W}$-algebra of the Kazama-Suzuki coset with $M>1$ cannot be given by a Drinfel'd-Sokolov reduction.
In fact, this property makes it very different from all previously considered cosets \cite{Gaberdiel:2010pz, Ahn:2011pv, Gaberdiel:2011nt, Creutzig:2011fe, Creutzig:2012ar, Beccaria:2013wqa} --- a fact which was overlooked in \cite{Creutzig:2013tja}.

\subsection{Dual higher spin theory}

Here we take the  duality between the $\shs_M[\lambda]$ Vasiliev theory of sec.~\ref{sec:matV} and the 't~Hooft limit of the coset~\eqref{eq:lcoset} as given.
From sec.~\ref{sec:def_KS} and \ref{sec:hsKS} we have learned that on the cylinder the Kazama-Suzuki coset~\eqref{eq:ks} is equivalent to the coset~\eqref{eq:lcoset} subject to the constraints~\eqref{eq:constr_KS}.
These constraints have an obvious  analogue on the higher spin side because they act only on the  currents of the coset and the correspondence between the (higher spin) currents of the coset theory and the higher spin theory is fully understood, compare eq.~\eqref{eq:gfix} with eqs.~(\ref{eq:coset_s1c}, \ref{eq:spin32_coset}, \ref{suSpins}).
Moreover, on the higher spin side these  constraints can be absorbed into the boundary conditions for the gauge fields, because according to section \ref{sec:asa} the asymptotic behavior of the latter is determined precisely by the higher spin currents.
Thus, in conclusion, the higher spin dual of the Kazama-Suzuki cosets is given again by the $\shs_M[\lambda]$ Vasiliev theory, but now the asymptotic boundary conditions include on top of eq.~\eqref{eq:bc} also the constraints~\eqref{eq:constr_KS}.


\section{Conclusion}


In this paper we have reconsidered the holographic dualities proposed in \cite{Gaberdiel:2013vva, Creutzig:2013tja} between the $\shs_M[\lambda]$ Vasiliev theory on $AdS_3$ and the 't~Hooft limit of the cosets~\eqref{eq:lcoset}
and~\eqref{eq:ks}. We have provided a  simplified proof for the agreement of the partition functions  and, in addition, have shown that it is the
asymptotic boundary conditions that determine which of the two cosets~\eqref{eq:lcoset}
or~\eqref{eq:ks} is the dual theory. In particular, this means that the number of superconformal symmetries in the Vasiliev theory depends on the chosen boundary conditions.

Let us now come back to the issues related to supersymmetry mentioned in the introduction. First, recall that
the cosets~\eqref{eq:lcoset} are supersymmetric for $M=1$, when  they reduce to the $\mathbb{C}P^{N}$   Kazama-Suzuki type cosets \cite{Kazama:1988qp}  which have
$\mathcal{N}=2$ superconformal symmetry, and for $M=2$, when  they correspond to the construction of \cite{Spindel:1988sr, VanProeyen:1989me, sevrin} based on Wolf spaces which guarantees non-linear large $\mathcal{N}=4$ superconformal symmetry \cite{Goddard:1988wv}.
However, we have explicitly checked that for $M>2$ the coset currents of spin $s=\frac{3}{2}$ do not generate any of the superconformal algebras classified in \cite{Fradkin:1992bz, Fradkin:1992km}.
The most one can do is to find four supercharges that generate the non-linear $\mathcal{N}=4$ superconformal algebra up to $1/c$ corrections proportional to bilinear terms in the spin $s=1$ currents.
One might hope that the situation improves in the 't~Hooft limit, where the central charge diverges and, naively, the problematic terms proportional to $1/c$ go away.
However, at the level of $\mathcal{W}$-algebras,  taking the 't~Hooft limit is not the same thing as letting $c \rightarrow \infty$.
One must also  rescale the currents by appropriate powers of $c$ so that the 't~Hooft limit becomes a classical limit \cite{Ahn:2012fz, Candu:2013uya};
only then can the quantum $\mathcal{W}$-algebra of the coset reproduce in the 't~Hooft limit the classical $\mathcal{W}$-algebra of asymptotic symmetries of the dual higher spin theory.
If the 't~Hooft limit is taken correctly, then the problematic terms proportional to $1/c$ survive.
In conclusion, the large $\mathcal{N}=4$ superconformal symmetry of the cosets~\eqref{eq:lcoset} is broken by $1/c$ corrections even in the 't~Hooft limit.



It would be interesting to check whether the large $\mathcal{N}=4$ superconformal symmetry of the extended Vasiliev theory is also broken by $1/c$ corrections.
In other words, the question is whether the 
non-linear large $\mathcal{N}=4$ superconformal algebra is a subalgebra of the Drinfel'd-Sokolov reduction of $\shs_M[\mu]$ for $M=2$ and whether it ceases to be a subalgebra for $M>2$.
The duality of \cite{Creutzig:2013tja} between the extended Vasiliev theory and the cosets~\eqref{eq:lcoset} predicts that this is indeed the case.
Thus, in order to put the duality on solid grounds, one should first confirm that the 
large $\mathcal{N}=4$ superconformal algebra is a subalgebra of the Drinfel'd-Sokolov reduction of $\shs_M[\mu]$ only for $M=2$
by a direct asymptotic symmetry analysis and then carry out a stronger check of the agreement between the asymptotic symmetry algebra of the higher spin theory and the coset $\mathcal{W}$-algebra along the lines of
\cite{Gaberdiel:2012ku, 
Candu:2012tr, Candu:2012ne,  Beccaria:2013wqa, Candu:2013uya} or \cite{Hanaki:2012yf, Ahn:2012fz, Ahn:2012vs}.

\acknowledgments{We thank Matthias Gaberdiel for numerous useful discussions and guidance.}

\appendix

\section{Coset currents}

In the basis \eqref{eq:dec_sunm} the OPEs of the current algebra $\SU(N+M)_{k}$ can be written as
\begin{align}\label{eq:sumn_OPE}
J^A(z)J^B(w)&\sim \frac{k \delta_{AB} }{(z-w)^2}+\frac{f_{ABC}J^C(w)}{z-w}\ ,\\ \notag
J^I(z)J^J(w)&\sim \frac{k \delta_{IJ} }{(z-w)^2}+\frac{{f_{IJK}} J^K(w)}{z-w}\ ,\;
 J(z)J(w)\sim \frac{kNM(N+M)}{(z-w)^2}\ ,\\ \notag
J^A(z)J^{ai}(w)&\sim \frac{t^A_{ba}J^{bi}(w)}{z-w}\ ,\; J^I(z)J^{ai}(w)\sim \frac{-t^I_{ij}J^{aj}(w)}{z-w}\ ,\; J(z)J^{ai}(w)\sim \frac{(N+M)J^{ai}(w)}{z-w}\ ,\\ \notag
J^A(z)\bar J^{ai}(w)&\sim \frac{-t^A_{ab}\bar J^{bi}(w)}{z-w}\ ,\;
J^I(z)\bar J^{ai}(w)\sim \frac{t^I_{ji}\bar J^{aj}(w)}{z-w}\ ,\;  J(z)\bar J^{ai}(w)\sim \frac{-(N+M)\bar J^{ai}(w)}{z-w}\ ,\\ \notag
 J^{ai}(z)\bar{J}^{bj}(w)&\sim \frac{k \delta_{ij}\delta_{ab}}{(z-w)^2} + \frac{\delta_{ij}t^A_{ba}J^A(w)-\delta_{ab}t^I_{ij}J^I(w)+\frac{1}{NM}\delta_{ij}\delta_{ab}J(w)}{z-w}\ ,
\end{align}
where $[t^A,t^B]={f_{ABC}} t^C$ is a basis of $\su(N)$, $[t^I,t^J]={f_{IJK}} t^K$ is a basis of $\su(M)$ and $t^A_{ab}$, $t^I_{ij}$ are their matrix elements in the fundamental representation.
We have for simplicity chosen these bases to be orthonormal, i.e.\ set $\tr t^A t^B = \delta_{AB}$ and $\tr t^I t^J = \delta_{IJ}$, so that $f_{ABC}$ and $f_{IJK}$ are completely antisymmetric.

The OPEs of the ``fermionic'' currents~\eqref{eq:twosuM}  with each other and with the fermions $\psi^{ai}$, $\bar \psi^{ai}$ have the form:
\begin{align}\label{eq:sumn_OPE_k}
K^A(z)K^B(w)&\sim \frac{M \delta_{AB} }{(z-w)^2}+\frac{{f_{ABC}}K^C(w)}{z-w}\ ,\\ \notag
K^I(z)K^J(w)&\sim \frac{N \delta_{IJ} }{(z-w)^2}+\frac{{f_{IJK}} K^K(w)}{z-w}\ ,\; K(z)K(w)\sim \frac{NM}{(z-w)^2}\ ,\\ \notag
K^A(z)\psi^{ai}(w)&\sim \frac{t^A_{ba}\psi^{bi}(w)}{z-w}\ ,\; K^I(z)\psi^{ai}(w)\sim \frac{-t^I_{ij}\psi^{aj}(w)}{z-w}\ ,\;K(z)\psi^{ai}(w)\sim \frac{ \psi^{ai}(w)}{z-w}\ ,\\ \notag
K^A(z)\bar \psi^{ai}(w)&\sim \frac{-t^A_{ab}\bar \psi^{bi}(w)}{z-w}\ ,\; K^I(z)\bar \psi^{ai}(w)\sim \frac{t^I_{ji}\bar \psi^{aj}(w)}{z-w}\ ,\;  K(z)\bar \psi^{ai}(w)\sim \frac{-\bar \psi^{ai}(w)}{z-w}\ .
\end{align}

The energy-momentum tensor of the coset~\eqref{eq:lcoset} is explicitly given by 
\begin{multline}
T = \frac{1}{2(k+N+M)} \left[(J^IJ^I) + (K^I K^I) +(J^{ai}\bar J^{ai})+(\bar J^{ai}J^{ai})- 2JK/NM -{}\right.\\
\left.{}-2K^A J^A-k :(\psi^{ai}\partial\bar \psi^{ai}+\bar \psi^{ai}\partial \psi^{ai}):\right]\ ,
\label{eq:cosetT_expl}
\end{multline}
which according to the Goddard-Kent-Olive construction \cite{Goddard:1986ee, god} can be computed as the difference between the energy momentum tensor of the numerator and the denominator.


\begin{thebibliography}{99}


\bibitem{Chang:2012kt}
  C.~-M.~Chang, S.~Minwalla, T.~Sharma and X.~Yin,
  ``ABJ Triality: from Higher Spin Fields to Strings,''
  J.\ Phys.\ A {\bf 46} (2013) 214009
  [arXiv:1207.4485 [hep-th]].




\bibitem{Aharony:2008ug}
  O.~Aharony, O.~Bergman, D.~L.~Jafferis and J.~Maldacena,
  ``${\cal N}=6$ superconformal Chern-Simons-matter theories, M2-branes and their gravity duals,''
  JHEP {\bf 0810}, 091 (2008)
  [arXiv:0806.1218 [hep-th]].


\bibitem{Aharony:2008gk}
  O.~Aharony, O.~Bergman and D.~L.~Jafferis,
  ``Fractional M2-branes,''
  JHEP {\bf 0811}, 043 (2008)
  [arXiv:0807.4924 [hep-th]].


\bibitem{Klebanov:2002ja}
  I.~R.~Klebanov and A.~M.~Polyakov,
  ``AdS dual of the critical O(N) vector model,''
  Phys.\ Lett.\ B {\bf 550} (2002) 213
  [hep-th/0210114].
	

\bibitem{Giombi:2009wh}
  S.~Giombi and X.~Yin,
  ``Higher Spin Gauge Theory and Holography: The Three-Point Functions,''
  JHEP {\bf 1009} (2010) 115
  [arXiv:0912.3462 [hep-th]].



\bibitem{Giombi:2010vg}
  S.~Giombi and X.~Yin,
  ``Higher Spins in AdS and Twistorial Holography,''
  JHEP {\bf 1104} (2011) 086
  [arXiv:1004.3736 [hep-th]].








\bibitem{Gaberdiel:2010pz}
  M.~R.~Gaberdiel and R.~Gopakumar,
  ``An AdS$_3$ Dual for Minimal Model CFTs,''
  Phys.\ Rev.\ D {\bf 83} (2011) 066007
  [arXiv:1011.2986 [hep-th]].


\bibitem{Gaberdiel:2012ku}
  M.~R.~Gaberdiel and R.~Gopakumar,
  ``Triality in Minimal Model Holography,''
  JHEP {\bf 1207} (2012) 127
  [arXiv:1205.2472 [hep-th]].


\bibitem{Gaberdiel:2012uj}
M.R.~Gaberdiel and R.~Gopakumar,
``Minimal model holography,''
J.\ Phys.\ A: Math.\ Theor.\ {\bf 46} (2013) 214002
{\tt [arXiv:1207.6697 [hep-th]]}.





\bibitem{Ahn:2011pv}
  C.~Ahn,
  ``The Large N 't~Hooft Limit of Coset Minimal Models,''
  JHEP {\bf 1110} (2011) 125
  [arXiv:1106.0351 [hep-th]].


\bibitem{Gaberdiel:2011nt}
  M.~R.~Gaberdiel and C.~Vollenweider,
  ``Minimal Model Holography for SO(2N),''
  JHEP {\bf 1108} (2011) 104
  [arXiv:1106.2634 [hep-th]].


\bibitem{Creutzig:2011fe}
  T.~Creutzig, Y.~Hikida and P.~B.~Ronne,
  ``Higher spin AdS$_3$ supergravity and its dual CFT,''
  JHEP {\bf 1202} (2012) 109
  [arXiv:1111.2139 [hep-th]].
	
\bibitem{Creutzig:2012ar}
  T.~Creutzig, Y.~Hikida and P.~B.~Ronne,
  ``N=1 supersymmetric higher spin holography on AdS$_3$,''
  JHEP {\bf 1302} (2013) 019
  [arXiv:1209.5404 [hep-th]].
	
\bibitem{Beccaria:2013wqa}
  M.~Beccaria, C.~Candu, M.~R.~Gaberdiel and M.~Groher,
  ``$\mathcal{N}=1$ extension of minimal model holography,'' JHEP {\bf 1307} (2013) 174
  [arXiv:1305.1048 [hep-th]].



\bibitem{Prokushkin:1998bq}
  S.~F.~Prokushkin and M.~A.~Vasiliev,
  ``Higher spin gauge interactions for massive matter fields in 3-D AdS space-time,''
  Nucl.\ Phys.\ B {\bf 545} (1999) 385
  [hep-th/9806236].


\bibitem{Gaberdiel:2013vva}
  M.~R.~Gaberdiel and R.~Gopakumar,
  ``Large $\mathcal{N}=4$ Holography,''
  JHEP {\bf 1309} (2013) 036
  [arXiv:1305.4181 [hep-th]].

\bibitem{Creutzig:2013tja}
  T.~Creutzig, Y.~Hikida and P.~B.~Ronne,
  ``Extended higher spin holography and Grassmannian models,''
  JHEP {\bf 1311} (2013) 038
  [arXiv:1306.0466 [hep-th]].


\bibitem{Candu:2012jq}
  C.~Candu and M.~R.~Gaberdiel,
  ``Supersymmetric holography on $AdS_3$,''
  JHEP {\bf 1309} (2013) 071
  [arXiv:1203.1939 [hep-th]].




\bibitem{deBoer:1999rh}
  J.~de Boer, A.~Pasquinucci and K.~Skenderis,
  ``AdS / CFT dualities involving large 2-D N=4 superconformal symmetry,''
  Adv.\ Theor.\ Math.\ Phys.\  {\bf 3} (1999) 577
  [hep-th/9904073].

\bibitem{Gukov:2004ym}
S.~Gukov, E.~Martinec, G.W.~Moore and A.~Strominger,
``The search for a holographic dual to AdS$_3 \times {\rm S}^{3}\times {\rm S}^{3} \times {\rm S}^1$,''
 Adv.\ Theor.\ Math.\ Phys.\  {\bf 9} (2005) 435
{\tt [arXiv:hep-th/0403090]}.


\bibitem{Spindel:1988sr} 
P.~Spindel, A.~Sevrin, W.~Troost and A.~Van Proeyen,
``Extended supersymmetric sigma models on group manifolds. 1. The complex structures,''
Nucl.\ Phys.\ B {\bf 308} (1988)  662.

\bibitem{VanProeyen:1989me} 
A.~Van Proeyen,
``Realizations of N=4 superconformal algebras on Wolf spaces,''
Class.\ Quant.\ Grav.\  {\bf 6} (1989) 1501.


\bibitem{sevrin}
  A.~Sevrin and G.~Theodoridis,
  ``N=4 Superconformal Coset Theories,''
  Nucl.\ Phys.\ B {\bf 332} (1990) 380.
	



	
	
\bibitem{Pope:1989sr}
  C.~N.~Pope, L.~J.~Romans and X.~Shen,
  ``$W(\infty)$ and the Racah-wigner Algebra,''
  Nucl.\ Phys.\ B {\bf 339} (1990) 191.



\bibitem{Vasiliev:1989re}
  M.~A.~Vasiliev,
  ``Higher Spin Algebras and Quantization on the Sphere and Hyperboloid,''
  Int.\ J.\ Mod.\ Phys.\ A {\bf 6} (1991) 1115.

	
	
\bibitem{Fradkin:1990qk}
  E.~S.~Fradkin and V.~Y.~Linetsky,
  ``Supersymmetric Racah basis, family of infinite dimensional superalgebras, $\SU(\infty + 1|\infty)$ and related 2-D models,''
  Mod.\ Phys.\ Lett.\ A {\bf 6} (1991) 617.



\bibitem{Ammon:2011ua}
  M.~Ammon, P.~Kraus and E.~Perlmutter,
  ``Scalar fields and three-point functions in D=3 higher spin gravity,''
  JHEP {\bf 1207} (2012) 113
  [arXiv:1111.3926 [hep-th]].


	\bibitem{Achucarro:1987vz}
  A.~Achucarro and P.~K.~Townsend,
  ``A Chern-Simons Action for Three-Dimensional anti-De Sitter Supergravity Theories,''
  Phys.\ Lett.\ B {\bf 180} (1986) 89.
\bibitem{Witten:1988hc}
  E.~Witten,
  ``(2+1)-Dimensional Gravity as an Exactly Soluble System,''
  Nucl.\ Phys.\ B {\bf 311} (1988) 46.




\bibitem{Vasiliev:2003ev}
  M.~A.~Vasiliev,
  ``Nonlinear equations for symmetric massless higher spin fields in (A)dS(d),''
  Phys.\ Lett.\ B {\bf 567} (2003) 139
  [hep-th/0304049].


	
\bibitem{Campoleoni:2010zq}
  A.~Campoleoni, S.~Fredenhagen, S.~Pfenninger and S.~Theisen,
  ``Asymptotic symmetries of three-dimensional gravity coupled to higher-spin fields,''
  JHEP {\bf 1011} (2010) 007
  [arXiv:1008.4744 [hep-th]].
	
\bibitem{Campoleoni:2012hp}
  A.~Campoleoni, S.~Fredenhagen, S.~Pfenninger and S.~Theisen,
  ``Towards metric-like higher-spin gauge theories in three dimensions,''
  J.\ Phys.\ A {\bf 46} (2013) 214017
  [arXiv:1208.1851 [hep-th]].
	
\bibitem{Gaberdiel:2011wb}
  M.~R.~Gaberdiel and T.~Hartman,
  ``Symmetries of Holographic Minimal Models,''
  JHEP {\bf 1105} (2011) 031
  [arXiv:1101.2910 [hep-th]].


\bibitem{Giombi:2008vd}
S.~Giombi, A.~Maloney, and X.~Yin, 
``One-loop partition functions of 3d gravity,'' 
JHEP {\bf 0808} (2008) 007
{\tt [arXiv:0804.1773 [hep-th]]}.

\bibitem{Gaberdiel:2010ar}
M.R.~Gaberdiel, R.~Gopakumar, and A.~Saha, 
``Quantum $W$-symmetry in AdS$_3$,''
JHEP {\bf 1102} (2011) 004
{\tt [arXiv:1009.6087 [hep-th]]}.
	
\bibitem{David:2009xg}
  J.~R.~David, M.~R.~Gaberdiel and R.~Gopakumar,
  ``The Heat Kernel on AdS(3) and its Applications,''
  JHEP {\bf 1004} (2010) 125
  [arXiv:0911.5085 [hep-th]].
		

\bibitem{Perlmutter:2012ds}
  E.~Perlmutter, T.~Prochazka and J.~Raeymaekers,
  ``The semiclassical limit of $\mathcal{W}_N$ CFTs and Vasiliev theory,''
  JHEP {\bf 1305} (2013) 007
  [arXiv:1210.8452 [hep-th]].
	
\bibitem{Hikida:2012eu}
  Y.~Hikida,
  ``Conical defects and $N=2$ higher spin holography,''
  JHEP {\bf 1308} (2013) 127
  [arXiv:1212.4124 [hep-th]].
	
	
\bibitem{god}
  P.~Goddard, A.~Kent and D.~I.~Olive,
  ``Virasoro Algebras and Coset Space Models,''
  Phys.\ Lett.\ B {\bf 152} (1985) 88.



\bibitem{Goddard:1986ee}
  P.~Goddard, A.~Kent and D.~I.~Olive,
  ``Unitary Representations of the Virasoro and Supervirasoro Algebras,''
  Commun.\ Math.\ Phys.\  {\bf 103} (1986) 105.

	
	
\bibitem{Lerche:1989uy}
  W.~Lerche, C.~Vafa and N.~P.~Warner,
  ``Chiral Rings in $\mathcal{N}=2$ Superconformal Theories,''
  Nucl.\ Phys.\ B {\bf 324} (1989) 427.


\bibitem{King:1971rs}
  R.~C.~King,
  ``Modification rules and products of irreducible representations of the unitary, orthogonal, and symplectic groups,''
  J.\ Math.\ Phys.\  {\bf 12} (1971) 1588.

\bibitem{King:1975vf}
  R.~C.~King,
  ``Branching Rules for Classical Lie Groups Using Tensor and Spinor Methods,''
  J.\ Phys.\ A {\bf 8} (1975) 429.









\bibitem{Gaberdiel:2011zw}
  M.~R.~Gaberdiel, R.~Gopakumar, T.~Hartman and S.~Raju,
  ``Partition Functions of Holographic Minimal Models,''
  JHEP {\bf 1108} (2011) 077
  [arXiv:1106.1897 [hep-th]].

\bibitem{Bouwknegt:1992wg}
  P.~Bouwknegt and K.~Schoutens,
  ``W symmetry in conformal field theory,''
  Phys.\ Rept.\  {\bf 223} (1993) 183
  [hep-th/9210010].

\bibitem{Ahn:2013bhw}
  C.~Ahn and J.~Paeng,
  ``Higher Spin Currents in the Holographic N=1 Coset Minimal Model,''
  arXiv:1310.6185 [hep-th].
	

\bibitem{deBoer:1993gd}
  J.~de Boer, L.~Feher and A.~Honecker,
  ``A Class of W algebras with infinitely generated classical limit,''
  Nucl.\ Phys.\ B {\bf 420} (1994) 409
  [hep-th/9312049].
	

\bibitem{Kazama:1988qp}
  Y.~Kazama and H.~Suzuki,
  ``New $\mathcal{N}=2$ Superconformal Field Theories and Superstring Compactification,''
  Nucl.\ Phys.\ B {\bf 321} (1989) 232.

	


\bibitem{Candu:2012tr}
  C.~Candu and M.~R.~Gaberdiel,
  ``Duality in N=2 Minimal Model Holography,''
  JHEP {\bf 1302} (2013) 070
  [arXiv:1207.6646 [hep-th]].
	

\bibitem{Candu:2012ne}
  C.~Candu, M.~R.~Gaberdiel, M.~Kelm and C.~Vollenweider,
  ``Even spin minimal model holography,''
  JHEP {\bf 1301} (2013) 185
  [arXiv:1211.3113 [hep-th]].



\bibitem{Hanaki:2012yf}
  K.~Hanaki and C.~Peng,
  ``Symmetries of Holographic Super-Minimal Models,''
  JHEP {\bf 1308} (2013) 030
  [arXiv:1203.5768 [hep-th]].



	
\bibitem{Candu:2013uya}
  C.~Candu and C.~Vollenweider,
  ``The $\mathcal{N} =$ 1 algebra $\mathcal{W}_\infty[\mu]$ and its truncations,''
  JHEP {\bf 1311} (2013) 032
  [arXiv:1305.0013 [hep-th]].
	

\bibitem{Ahn:2012fz}
  C.~Ahn,
  ``The Large N 't~Hooft Limit of Kazama-Suzuki Model,''
  JHEP {\bf 1208} (2012) 047
  [arXiv:1206.0054 [hep-th]].


\bibitem{weyl}
H.~Weyl,
 The Classical Groups: Their Invariants and Representations,
Princeton University Press (1939).




\bibitem{Goddard:1988wv}
  P.~Goddard and A.~Schwimmer,
  ``Factoring Out Free Fermions and Superconformal Algebras,''
  Phys.\ Lett.\ B {\bf 214} (1988) 209.


	
\bibitem{Fradkin:1992bz}
  E.~S.~Fradkin and V.~Y.~Linetsky,
  ``Results of the classification of superconformal algebras in two-dimensions,''
  Phys.\ Lett.\ B {\bf 282} (1992) 352
  [hep-th/9203045].
	
\bibitem{Fradkin:1992km}
  E.~S.~Fradkin and V.~Y.~Linetsky,
  ``Classification of superconformal and quasisuperconformal algebras in two-dimensions,''
  Phys.\ Lett.\ B {\bf 291} (1992) 71.
	

\bibitem{Henneaux:1999ib}
  M.~Henneaux, L.~Maoz and A.~Schwimmer,
  ``Asymptotic dynamics and asymptotic symmetries of three-dimensional extended AdS supergravity,''
  Annals Phys.\  {\bf 282} (2000) 31
  [hep-th/9910013].
		
\bibitem{Ahn:2012vs}
  C.~Ahn,
  ``The Operator Product Expansion of the Lowest Higher Spin Current at Finite N,''
  JHEP {\bf 1301} (2013) 041
  [arXiv:1208.0058 [hep-th]].
		
\end{thebibliography}
\end{document}